\documentclass[aps,pre,epsf,superscriptaddress,amsmath,amssymb,amsfonts,showpacs,superscriptaddress,floatfix,preprintnumbers,twocolumn,nofootinbib,10pt]{revtex4-1}
\usepackage{microtype}
\DisableLigatures[f]{encoding = *, family = *}
\usepackage{graphicx}
\usepackage{epsfig}
\usepackage{dcolumn}
\usepackage{bm}
\usepackage{braket}
\usepackage{amsmath}
\usepackage{mathtools}
\usepackage{color}
\newcommand{\abs}[1]{\left| #1 \right|}
\newcommand{\iu}{{i\mkern1mu}}

\begin{document}
	
\title{Stationary and dynamical properties of two harmonically trapped bosons in the crossover from two dimensions to one}
	
\author{G. Bougas}
\affiliation{Center for Optical Quantum Technologies, Department of Physics, University of Hamburg, 
Luruper Chaussee 149, 22761 Hamburg Germany}
\author{S. I. Mistakidis}
\affiliation{Center for Optical Quantum Technologies, Department of Physics, University of Hamburg, 
Luruper Chaussee 149, 22761 Hamburg Germany} 
\author{G. M. Alshalan}
\affiliation{ Department of Physics and Research Laboratory of Electronics, Massachusetts Institute of Technology, Cambridge, Massachusetts 02139, USA}
\author{P. Schmelcher}
\affiliation{Center for Optical Quantum Technologies, Department of Physics, University of Hamburg, Luruper Chaussee 149, 22761 Hamburg Germany} 
\affiliation{The Hamburg Centre for Ultrafast Imaging, Universit\"at Hamburg Luruper Chaussee 149, 22761 Hamburg, Germany}
	
\date{\today}

\begin{abstract}
We unravel the stationary properties and the interaction quench dynamics of two bosons, confined in a two-dimensional anisotropic harmonic trap. 
A transcendental equation is derived giving access to the energy spectrum and revealing the dependence of the energy gaps on the anisotropy parameter. 
The relation between the two and the one dimensional scattering lengths as well as the Tan contacts is established. 
The contact, capturing the two-body short range correlations, shows an increasing tendency for a larger anisotropy. 
Subsequently, the interaction quench dynamics from attractive to repulsive values and vice versa is investigated for various anisotropies. 
A closed analytical form of the expansion coefficients of the two-body wavefunction, during the time evolution is constructed. 
The response of the system is studied by means of the time-averaged fidelity, the spectra of the spatial extent of the cloud in each direction and the one-body density. 
It is found that as the anisotropy increases, the system becomes less perturbed independently of the interactions while for fixed anisotropy quenches towards the 
non-interacting regime perturb the system in the most efficient manner. 
Furthermore, we identify that in the tightly confined direction more frequencies are involved in the dynamics stemming from higher-lying excited states.
\end{abstract}

\maketitle

\section{Introduction}

Ultracold gases offer a highly controllable platform for studying quantum few- and many-body systems due to their extraordinary tunability \cite{Bloch,Blume}. 
Feshbach resonances play a pivotal role, since the interparticle interaction strength can be arbitrarily adjusted by means of magnetic and optical fields \cite{Fesh1,Fesh2}. 
Moreover, advances in atom trapping enable us to realize systems of different dimensionality \cite{Petrov,lower-D,Merlotti,Boettcher} and particle number, thus rendering 
few-body ensembles which exhibit remarkable properties, such as the Efimov effect, experimentally tractable \cite{Brouzos,Lompe,Blume,Sowinski,Greene,Jochim}.  

Utilizing anisotropic harmonic traps allows to reach the quasi two-dimensional (quasi-2D) and quasi-one-dimensional (quasi-1D) regimes by manipulating the axial ($\omega_z$) or the 
radial frequency ($\omega_r$), such that $\hbar \omega_z$ ($\hbar \omega_r$) becomes much larger than all the intrinsic energy scales of the system \cite{He,Boettcher,Lia}. 
The crossover to different dimensions has been investigated in various setups and several relations have been established for the scattering properties in different dimensions, e.g. 
between the scattering lengths \cite{Crossover1,Crossover2,Crossover3,Wetterich,Crossover5,Wetterich2,Crossover6}. 
These relations give rise to confinement induced resonances \cite{Crossover1,CIR,CIR_exp,Vladimir,Giannakeas}, which provide further means to tune the interparticle interaction in lower dimensional settings. 
Moreover, it has been showcased that the two-body Tan contact in three-dimensions (3D) and in 2D, 1D are proportional by factors depending on the dimension \cite{He,Decamp,Valiente}. 
Interestingly, next-to-leading order terms in the asymptotic expansion of the two-body momentum distribution reveal the contribution of the three-body contact and the role of dimensionality \cite{Bellotti}. 
Recently, on the two-body level, a correspondence between a dimension dependent centrifugal barrier and a confining potential has been established \cite{Jensen}. 
Importantly, apart from the stationary properties, the non-equilibrium dynamics of Bose and Fermi gases at the dimensional crossover has attracted considerable interest \cite{Peppler,Holten,Kottke,Kronke}. 
This is corroborated by the advent of new trapping techniques e.g. utilizing optical tweezers \cite{Liu,Doyle} which paves the way for monitoring the time-evolution of few-body systems. 
For instance, the collisional dynamics of two $^6$Li atoms \cite{Jochim} has been experimentally probed, by quenching the frequencies of an anisotropic 3D harmonic oscillator. 

The stationary properties of two ultracold atoms confined in an isotropic harmonic oscillator trap have been thoroughly explored across all dimensions \cite{Busch,Zyl,Shea}. 
Generalizations of these studies include for instance the involvement of anisotropic traps in three dimensions \cite{Calarco,Calarco2,Bolda,Chen}, higher partial waves \cite{Stock,Zinner2}, long-range 
interactions \cite{Koscic} and hard-core interaction potentials \cite{Diakonos}. 
Moreover, a correspondence between three bosons interacting via three body forces in 1D and two bosons interacting via pairwise interactions in 2D has been established \cite{Valiente1,Valiente2,Sekino,Nishida,Guijarro,Pricoupenko}. 
The stationary solutions have been utilized in order to probe the non-equilibrium dynamics of two atoms, by quenching the interaction strength in all dimensions \cite{Bougas,Laura,Bolsinger,Fogarty}. The solutions also
serve as a simple model for the dynamics of quenched Bose gases, at short times and larger momenta than those set by the density of the gas \cite{Sykes,Corson2}. 
Analytical expressions for several observables are known, including for instance momentum distributions \cite{Corson} and thermodynamical quantities \cite{Garcia,Ikeda}. 

Even though the dimensional crossover at the two-body level has been extensively studied from three to lower dimensions, the crossover from two to one dimensions is yet an unexplored problem, in terms of both the stationary and the dynamical properties. 
In this work, we shed light into the stationary properties and interaction quench dynamics of two ultracold bosons trapped in an anisotropic 2D harmonic trap. However, our results have a more general character and can be equally applied to two distinguishable $s$-wave interacting ultracold atoms in even-parity states of their relative coordinate. 
A transcendental equation for the anisotropic system is derived allowing us to probe the underlying energy spectrum for arbitrary interactions and anisotropies. 
For instance, it is shown that the energy gaps between the involved eigenstates for a fixed interaction strength strongly depend on the anisotropy. An analytical expression for the two-boson wavefunction both in real and momentum space is constructed and the relation between the 2D and the 1D scattering 
lengths is established. 
We find that the momentum distribution exhibits a multihump structure along the weaker confined direction while the corresponding one-body densities feature two-hump patterns. 
Remarkably, the 2D and the 1D Tan contacts, capturing the occurrence of short-range two-body correlations, are found to be proportional to each other by a simple relation. 
The Tan contact of the bound and the ground state shows an increasing tendency for larger anisotropies independently of the sign of the interaction, and in particular for the ground 
state it tends to saturate when approaching the 1D regime. 

Subsequently, we focus on the interaction quench dynamics of the two particles from attractive to repulsive interactions and vice versa. 
The response of the system is analyzed in terms of the time-averaged fidelity, and the frequency spectra of the spatial extent of the bosonic cloud in both confined directions. 
We showcase that the time-evolved state deviates significantly from the initial one in the vicinity of zero postquench interactions, when the latter is initialized at finite attractive or repulsive interactions. For increasing anisotropy the system 
becomes less perturbed following an interaction quench, independently of the interactions. 
The quench excites a breathing motion, visualized in the time-evolution of the reduced one-body density, in both the $x$ and $y$ directions with a distinct number of 
participating frequencies in each spatial direction. 

This work is structured as follows. 
In Sec. \ref{Setup}, we introduce our setup of the two trapped bosons in a 2D anisotropic harmonic trap. 
Subsequently, in Sec. \ref{Energy Spectra} the energy spectra are presented for various anisotropies, while Sec. \ref{wavefunctions} contains the expression of the two-body 
wavefunction in real and momentum space. 
Section \ref{one_body_stat} is dedicated to the behavior of the reduced one-body density for several anisotropy parameters and Sec. \ref{Tan_stat} showcases the Tan contact of the bound and the 
ground states with respect to the anisotropy. 
In Sec. \ref{dynamics}, the interaction quench dynamics of two bosons is explored for different anisotropies. We lay out our concluding remarks and provide an outlook in Sec. \ref{conclusion}. 
Appendix \ref{Trans}, provides details on the derivation of the transcendental equation which determines the relative energy of the two bosons. 
Appendix \ref{Recovery} provides the 1D energy spectrum of two bosons by inspecting the quasi-1D limit of the transcendental equation. 
Details on the calculation of the 2D Tan contact and its quasi-1D limit are presented in appendix \ref{Ap:Tan}. 
Appendix \ref{Ap:Freq} includes an analytical derivation of the spatial extent of the bosonic cloud in both directions and the corresponding frequency amplitudes.

\section{Hamiltonian and eigenvalue problem} \label{Setup}
	
We consider two ultracold bosons trapped in a 2D anisotropic harmonic trap interacting via an $s$-wave pseudo-potential. 
Note that the following analysis applies to the general case of ultracold atoms except for two spin-polarized fermions \cite{Shea}, see in particular the discussion following Eq. (7). 
The latter constitutes an adequate approximation within the ultracold regime \cite{Fesh1,Fesh2}. 
The Hamiltonian of the system reads 
\begin{equation}
\mathcal{H}=\sum_{i=1}^2 \left[-\frac{\hbar^2}{2m} \bm{\nabla}_i^2 +m\omega_x^2\frac{(x_i^2+\alpha^2y_i^2)}{2}\right]+2V_{\textrm{pp}}\left(\bm{\rho}_1-\bm{\rho}_2\right). \label{Hamiltonian}
\end{equation} 
For simplicity, below, we shall adopt harmonic oscillator units namely $\hbar=m=\omega_x=1$ unless it is stated otherwise.  
Additionally, the anisotropy parameter $\alpha=\frac{\omega_y}{\omega_x}$ is the ratio of the harmonic trap frequencies along the $y$ and $x$ spatial directions.
Evidently, $\alpha$ takes values from unity (2D case) up to infinity (1D case). 
Also, $\bm{\rho}_i=(x_i,y_i)$ denotes the position of the $i$-th boson in the 2D plane whilst the prefactor 2 in Eq. (\ref{Hamiltonian}) is used for later convenience. 
The zero range regularized $s$-wave pseudo-potential assumes the following form \cite{Olshanii1}
\begin{equation}
V_{\textrm{pp}}(\bm{\rho})= -\frac{\pi \delta(\bm{\rho})}{\ln(Aa_{\textrm{2D}}\Lambda)}\left(1-\ln(A\Lambda \rho)\rho\frac{\partial}{\partial \rho}  \right),
\label{Pot}
\end{equation}
where $\Lambda$ is an arbitrary dimensionful parameter possessing the units of momentum and $A=e^{\gamma}/2$ with $\gamma=0.577\ldots$ being the Euler-Mascheroni constant. 
Note that the arbitrary parameter $\Lambda$ does not affect any observable of the system and eventually drops out of the calculations when the pseudo-potential is applied 
to wavefunctions exhibiting a logarithmic behavior at the origin $\rho=0$ \cite{Olshanii1,Olshanii2}. 
The 2D $s$-wave scattering length is $a_{2D}$. 
	
To separate the center-of-mass ($X$, $Y$) and relative ($x$, $y$) coordinates, we employ the following transformations in terms 
of the Cartesian coordinates ($x_i$, $y_i$) $X=\frac{x_1+x_2}{\sqrt{2}}, Y=\frac{y_1+y_2}{\sqrt{2}}$ and $x=\frac{x_1-x_2}{\sqrt{2}}, y=\frac{y_1-y_2}{\sqrt{2}}$.
Therefore, the Hamiltonian of Eq. (\ref{Hamiltonian}) separates into the center-of-mass $\mathcal{H}_{\textrm{c.m.}}$ and the relative $\mathcal{H}_{\textrm{rel}}$ Hamiltonian, 
namely $\mathcal{H}=\mathcal{H}_{\textrm{c.m.}}+\mathcal{H}_{\textrm{rel}}$ with 
\begin{eqnarray}
\mathcal{H}_{\textrm{c.m.}}&=& -\frac{1}{2} (\partial_X^2+\partial_Y^2)+\frac{1}{2}(X^2+\alpha^2 Y^2) \nonumber \\
\mathcal{H}_{\textrm{rel}} &=& -\frac{1}{2} (\partial_x^2+\partial_y^2)+\frac{1}{2}(x^2+\alpha^2 y^2)- \nonumber \\ & &
-\frac{\pi \delta(x)\delta(y)}{\ln(Aa_{\textrm{2D}}\Lambda)}\left[1-\ln(\sqrt{2}A\Lambda \rho)\rho \frac{\partial}{\partial \rho} \right],
\label{Hamilts}
\end{eqnarray}
where $\rho=\sqrt{x^2+y^2}$. Due to the above-described separation of the Hamiltonian, the corresponding wavefunction of the system can subsequently be written as a 
product state i.e. $\Psi(\boldsymbol{\rho}_1,\boldsymbol{\rho}_2)=\Psi_{\textrm{c.m.}}(X,Y)\Psi_{\textrm{rel}}(x,y)$. 

The eigenvalue problem of the center-of-mass is easy to solve since it consists of two decoupled non-interacting 1D harmonic oscillators in the $x$ and $y$ directions, see Eq. (\ref{Hamilts}). 
Indeed, the corresponding wavefunction reads
\begin{equation}
\Psi_{\textrm{c.m.}}(X,Y)=\phi_n(X)\phi_m(Y),
\label{Wave_CM}
\end{equation}  
where $\phi_n(z)=\frac{e^{-\omega z^2/2}}{\sqrt{2^nn!}}\left(\frac{\omega}{\pi}\right)^{1/4}H_n(\sqrt{\omega}z)$ with $n=0,1,2,\dots$ are the eigenfunctions of 
a 1D harmonic oscillator of frequency $\omega=1,\alpha$ and energy $E_n=(n+1/2)\omega$ in harmonic oscillator units \cite{Sakurai}. 
$H_n$ are the Hermite polynomials of degree $n$. 
Thus, the energy of the center-of-mass reads $E_{\textrm{c.m.}}^{\tilde{n},\tilde{m}}=\tilde{n}+\alpha \tilde{m} +\frac{\alpha+1}{2}$. 
Throughout this work, we assume that the center-of-mass wavefunction is in its ground state $\Psi_{\textrm{c.m.}}(X,Y)=\phi_0(X)\phi_0(Y)$. 

To tackle the eigenvalue problem of the relative Hamiltonian, $\mathcal{H}_{\textrm{rel}}$, we utilize as a wavefunction ansatz an expansion 
over the non-interacting eigenstates $\phi_n(z)$ \cite{Busch,Laura} in both spatial directions i.e. 
\begin{equation}
\Psi_{\textrm{rel}}(x,y)=\sum_{n,m} c_{n,m}\phi_n(x)\phi_m(y).
\label{ansatz}
\end{equation} 
Here, $c_{n,m}$ denote the corresponding expansion coefficients (see also below). 
By plugging Eq. \eqref{ansatz} into the Schr\"odinger equation for the relative Hamiltonian $\mathcal{H}_{\textrm{rel}}\Psi_{\textrm{rel}}=E_{\textrm{rel}}\Psi_{\textrm{rel}}$, 
see also Eq. \eqref{Hamilts}, and projecting onto the non-interacting eigenstates $\phi_{n'}^*(x)\phi_{m'}^*(y)$, one arrives at the following equation 
\begin{gather}
0=c_{n',m'}(E_{\textrm{rel}}^{n',m'}-E_{\textrm{rel}})  \nonumber \\-\frac{\pi \phi_{n'}^*(0)\phi_{m'}^*(0)}{\ln (a_{\textrm{2D}}A\Lambda)}\left\{  \left(1-\ln (\sqrt{2}A\Lambda\rho)\rho \frac{\partial}{\partial \rho} \right)\Psi_{\textrm{rel}}(x,y) \right\}_{\rho\rightarrow 0},
\label{coeffs_expression} 
\end{gather}
where $\rho=\sqrt{x^2+y^2}$ and $E_{\textrm{rel}}^{n,m}=n+\alpha m+\frac{\alpha+1}{2}$. 
The regularization operator enclosed in the parentheses $( \dots )$ of Eq. \eqref{coeffs_expression} acts on the relative wavefunction, and subtracts the logarithmic divergence 
close to the origin, $\rho=0$ \cite{Olshanii2,Wodkiewicz}. 
As a consequence, the expression in the right hand side of Eq. \eqref{coeffs_expression} is related to a normalization factor denoted below by $B$ of the wavefunction, as it 
has been argued in \cite{Busch,Laura}, that will be determined later. 
The expansion coefficients, $c_{n,m}$, thus take the following form
\begin{equation}
c_{n,m}=B\frac{\phi_n^*(0)\phi_m^*(0)}{E_{\textrm{rel}}^{n,m}-E_{\textrm{rel}}}.
\label{coeffs}
\end{equation} 
Note that the expansion coefficients vanish for odd $n,m$. 
Indeed, the 2D pseudo-potential of Eq. \eqref{Pot} affects only states with a non-vanishing value at $x=y=0$ which in turn involve 
only even Hermite polynomials, i.e. even-parity states of the relative coordinate, in the ansatz \eqref{ansatz} \cite{Calarco,Calarco2}. Therefore our analysis is also valid for two distinguishable ultracold atoms in even-parity states, i.e. the ones that are affected by the $s$-wave interaction. The odd-parity states are not impacted by the contact potential. Having at hand the expansion coefficients, see Eq. \eqref{coeffs}, one can directly perform the double summation appearing in Eq. \eqref{ansatz}.
For that end, we express the denominator of the expansion coefficients [Eq. \eqref{coeffs}] in an integral representation \cite{Busch,Calarco,Calarco2} 
\begin{equation}
\frac{1}{E_{\textrm{rel}}^{n,m}-E_{\textrm{rel}}}=\int_{0}^{\infty} dt\, e^{-t(E_{\textrm{rel}}^{n,m}-E_{\textrm{rel}})},
\label{Int_rep}
\end{equation}
and then perform the double summation by using the Mehler identity for the Hermite polynomials \cite{Ismail}. 
Therefore, the relative wavefunction reads 
\begin{gather}
\Psi_{\textrm{rel}}(x,y)=B \frac{\sqrt{\alpha}}{2\pi} e^{-(x^2+\alpha y^2)/2} \times \nonumber \\ \int_{0}^{+\infty} dt \, 
\exp\left( \frac{e^{-t}x^2}{e^{-t}-1}+\frac{\alpha e^{-\alpha t}y^2}{e^{-\alpha t}-1} \right) \frac{e^{-tf(E)/2}}{\sqrt{1-e^{-t}}\sqrt{1-e^{-\alpha t}}},
\label{wavefunction_integral}
\end{gather}
where $f(E)=\frac{\alpha+1}{2}-E$. 
The above integral converges provided that $f(E)>0$. 
Later on, and in particular in Appendix \ref{Trans}, we shall consider values of $f(E)<0$ by means of analytic continuation \cite{Calarco,Calarco2}. 
Note also that in Eq. (\ref{wavefunction_integral}) we have dropped the subscript $\textrm{rel}$ from the energy for simplicity. 

Furthermore, by employing the form of the expansion coefficients [Eq. \eqref{coeffs}] the relative energy is determined via 
Eq. \eqref{coeffs_expression}, namely 
\begin{equation}
\left \{  \left(1-\ln(\sqrt{2}A\Lambda \rho)\rho \frac{\partial}{\partial \rho}\right) \frac{\Psi_{\textrm{rel}}(x,y)}{B}  \right \}_{\rho\rightarrow 0}=\frac{\ln (a_{\textrm{2D}}A\Lambda)}{\pi},
\label{energy}
\end{equation} 
where $\Psi_{\textrm{rel}}(x,y)$ is determined by Eq. \eqref{wavefunction_integral}. The aim of the following section is to solve Eq. \eqref{energy} for an arbitrary anisotropy parameter $\alpha$, in order to determine the stationary 
properties of the two bosons by calculating their energy spectra and eigenstates.

\section{Energy Spectra} \label{Energy Spectra} 

\subsection{Transcendental equation}

To find the relative energy $E$ we need to solve Eq. \eqref{energy} and therefore establish a formula that captures the behavior of the wavefunction close to $x=y=0$. 
For $x,y\to 0$, the main contribution to the integral \eqref{wavefunction_integral} stems from very small values of the integration variable $t$ \cite{Calarco,Calarco2}.  
Indeed, the integral appearing in Eq. \eqref{wavefunction_integral} can be splitted into two parts
\begin{eqnarray}
\Psi_{\textrm{rel}}(x,y)|_{x,y\ll 1}&=& \frac{B}{2\pi} \int_{0}^{L} dt\, \frac{e^{-(x^2+y^2)/t}}{t} \nonumber \\& & +B\frac{\sqrt{\alpha}}{2\pi}\underbrace{\int_{L}^{+\infty} dt\, 
\frac{e^{-tf(E)/2}}{\sqrt{1-e^{-t}}\sqrt{1-e^{-\alpha t}}}}_{I(f(E)/2)}.
\label{integral_split} \nonumber \\
\end{eqnarray} 
In the first part, we have linearized all the exponentials around $t=0$ while in the second part we have set $x=y=0$ directly. 
The parameter $L$ is very small being of the order of $x,y$. 
The first integral corresponds to $\Gamma\left(0,\frac{x^2+y^2}{L}\right)$, where $\Gamma(x,y)$ is the incomplete gamma function \cite{Stegun}. 
For small $r^2=x^2+y^2$, this gamma function can be expanded as follows 
\begin{equation}
\Gamma\left(0,\frac{r^2}{L}\right) \stackrel{r\rightarrow 0}{\longrightarrow} -\gamma -\ln \left(\frac{r^2}{L}\right) + \frac{r^2}{L} +\mathcal{O}(r^4).
\label{gamma_expansion}
\end{equation}
Note that this result is independent of $\alpha$, since at very small interparticle distances $r\to0$ the confining potential does not play any crucial role and the wavefunction develops a logarithmic behavior, as a consequence of the 2D interaction pseudo-potential \cite{He,Makhalov}. 
At this point it is better to restore the units, i.e. $x^2+y^2\rightarrow \frac{x^2+y^2}{l_x^2}$, where $l_x=\sqrt{\frac{\hbar}{m\omega_x}}$ is the harmonic 
oscillator length in the $x$ direction. 
Thus, we can deduce that the pure 2D regime is accessed when the interparticle distance $r$ is much smaller than $l_x$. 

Since the behavior of the relative wavefunction $\Psi_{\textrm{rel}}(x,y)$ is now available (see Eq. \eqref{integral_split}) close to $x=y=0$, one can insert Eq. \eqref{integral_split} into Eq. \eqref{energy} and in turn 
derive a transcendental equation that will allow us to determine the relative energy of the two bosons [see for more details Appendix \ref{Trans}]. 
The resulting transcendental equation reads 
\begin{equation}
-\gamma +2\ln 2+\sqrt{\alpha}\underbrace{\int_{0}^{1} dz \, \ln(1-z)\varphi'\left(z,\frac{f(E)}{2}\right)}_{P(f(E)/2)}=-\frac{1}{g},
\label{energy_spectrum2}
\end{equation}
where $g=\left(\ln \left(\frac{1}{2a_{\textrm{2D}}^2}\right)\right)^{-1}$ is the 2D coupling constant \cite{Busch,Zyl,Doganov}, $\varphi(z,f(E)/2)=z^{f(E)/2-1}\frac{\sqrt{1-z}}{\sqrt{1-z^{\alpha}}}$, 
and the differentiation is performed with respect to the variable $z$. 
Eq. \eqref{energy_spectrum2} provides the energy spectrum of the two bosons for an arbitrary anisotropy parameter $\alpha$. 
As it has been mentioned earlier, this equation is valid only for $f(E)>0$. 
Its extension to negative values is granted by the recurrence formula [see also Appendix \ref{Trans}] 
\begin{eqnarray}
P\left(\frac{f(E)}{2}\right)&=&P\left(\alpha+\frac{f(E)}{2}\right)+ \nonumber \\& & \sum_{n=0}^{\infty} \binom{1/2}{n}\frac{\sqrt{\pi}(-1)^n\Gamma\left(
\frac{f(E)}{2}+\alpha n\right)}{\Gamma\left(\frac{1}{2}+\frac{f(E)}{2}+\alpha n\right)}. \nonumber \\
\label{num_continuation}
\end{eqnarray} 

\subsection{Quasi-1D limit}

Before calculating the energies for various values of $\alpha$, let us first retrieve the 1D energy spectrum, by assuming that $\alpha \gg 1$. 
In this case the harmonic confinement along the $y$ direction is tight and therefore we enter the quasi-1D regime, at least when the interparticle distance is 
comparable or larger than the harmonic oscillator length in the $x$ direction i.e. $r\geq l_x$ (see also the previous discussion). 
For $\alpha \gg 1$, the transcendental equation \eqref{energy_spectrum2} becomes [see also Appendix \ref{Recovery}] 
\begin{figure}[t!]
\centering
\includegraphics[width=0.47 \textwidth]{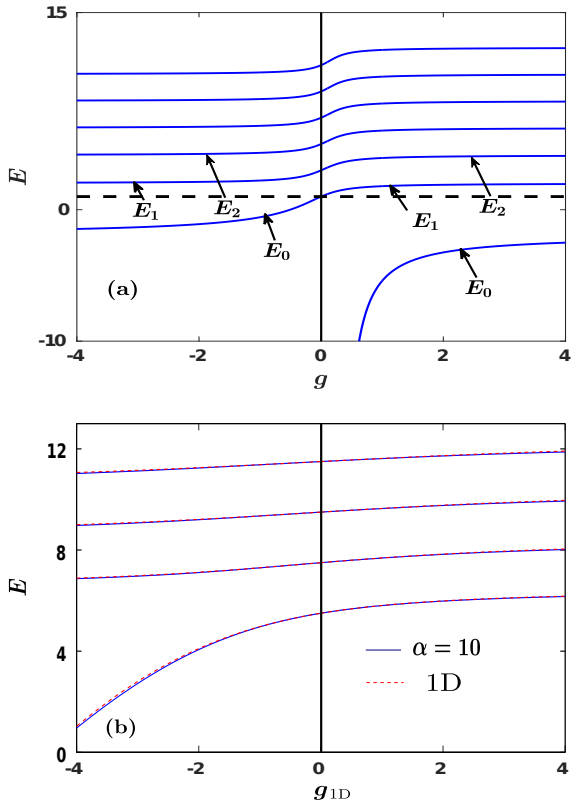}
\caption{(a) Energy spectrum with anisotropy $\alpha=1$, thus recovering the 2D limit, for various 2D interaction strengths. The black dashed line indicates the zero point energy. 
(b) Comparison of the energy spectra for $\alpha=10$ (blue line) and for a pure 1D system (red dashed line), with respect to the 1D interaction strength $g_{\textrm{1D}}$. 
In both cases the system consists of two ultracold bosons confined in an anisotropic 2D harmonic trap. All quantities shown are in dimensionless units.}
\label{Fig:Comparison}
\end{figure}

\begin{equation}
    \sqrt{\pi\alpha} \frac{\Gamma\left(\frac{f(E)}{2} \right)}{\Gamma\left(\frac{1}{2}+\frac{f(E)}{2} \right)}-\ln (\alpha) +D=\ln(a_{\textrm{2D}}^2), 
    \label{quasi_1D}
\end{equation}
where $D=-\gamma-2\sqrt{k}+\ln(2k)+\frac{k}{4}-\frac{k^2}{192}-\frac{k^3}{1152}$ and $k\approx 6$, see for details Appendix \ref{Recovery}. 
The above formula is reminiscent of the transcendental equation of two bosons confined in a 1D harmonic trap, which determines the energy spectrum of this system 
and reads \cite{Busch}
\begin{equation}
\sqrt{2}a_{\textrm{1D}}=\frac{\Gamma\left(\frac{1}{4}-\frac{E}{2} \right)}{\Gamma \left(\frac{3}{4}-\frac{E}{2} \right)}=-\frac{2\sqrt{2}}{g_{\textrm{1D}}}.
\label{spectrum_1D}
\end{equation} 
This expression is derived by following the same steps as in Section \ref{Setup} but in 1D and with the pseudo-potential 
$V_{\textrm{pp}}(x)=-\frac{2}{a_{\textrm{1D}}}\delta(x)$ \cite{CIR}. 
Most importantly, employing a proper rescaling of the energies in Eq. \eqref{quasi_1D}, namely $E'=-f(E)+1/2$ and comparing Eqs. \eqref{quasi_1D}, \eqref{spectrum_1D}, 
we obtain a relation between the 2D, $a_{\textrm{2D}}$, and the 1D, $a_{\textrm{1D}}$, scattering lengths 
\begin{figure*}[t!]
\centering
\includegraphics[width= 1 \textwidth]{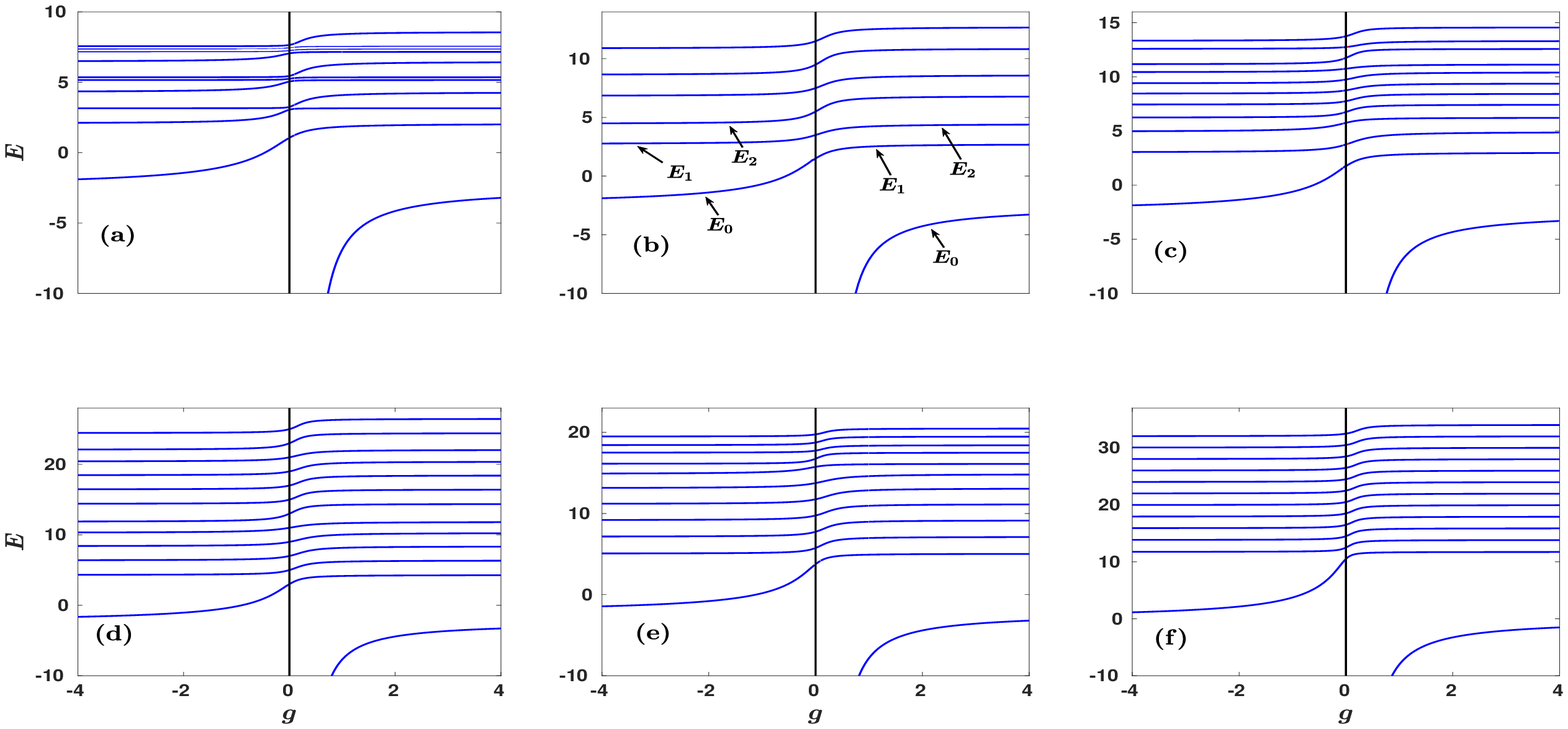}
\caption{Energy spectra for anisotropy parameter (a) $\alpha=1.1$, (b) $\alpha=2$, (c) $\alpha=2.5$, (d) $\alpha=5$, (e) $\alpha=6.5$ and (f) $\alpha=20$ for varying 2D coupling 
strength $g$. 
The labeling of the energy states is showcased only in panel (b) for convenience. In all cases the quantities displayed are in dimensionless units.}
\label{Fig:Spectra}
\end{figure*}

\begin{equation}
a_{\textrm{2D}}=\frac{D_0}{\sqrt{\alpha}}e^{\sqrt{\pi \alpha}a_{\textrm{1D}}/\sqrt{2}},
\label{Comparison}
\end{equation} 
with $D_0=e^{D/2}$. 
We remark that when restoring the units of the system, this expression acquires the form $a_{\textrm{2D}}=l_yD_0e^{\sqrt{\pi }a_{\textrm{1D}}/(\sqrt{2}l_y)}$, 
where $l_y$ is the harmonic oscillator length in the $y$ direction. 
Recently, a similar relation between these two scattering lengths has been established in Ref. \cite{Crossover5}, by means of non relativistic effective field theory. 
The connection between the scattering lengths is achieved by imposing periodic boundary conditions along one direction and comparing the effective range expansion 
with the purely 1D one. 
Apart from the scattering lengths, we are able to establish also a relation among the coupling constants in one 
and two dimensions
\begin{equation}
\frac{1}{g}=\ln (\alpha)-\ln 2+\frac{2\sqrt{2\pi \alpha}}{g_{\textrm{1D}}}-D,
\label{couplings}
\end{equation} 
where $g$ denotes the 2D effective coupling constant and $g_{\textrm{1D}}$ the corresponding 1D effective interaction strength, see also Eq. \eqref{spectrum_1D}. 

Let us also note in passing that the 2D energy spectrum can also be easily retrieved. 
Indeed, by substituting $\alpha=1$ in Eq. \eqref{wavefunction_integral}, one gets 
\begin{eqnarray}
\Psi_{\textrm{rel}}(x,y)&=&\frac{B}{2\pi} e^{-(x^2+y^2)/2}\nonumber  \\ & &\Gamma\left(\frac{f(E)}{2}\right)U\left(\frac{f(E)}{2},1,x^2+y^2\right),
\label{2D_wavefunction}
\end{eqnarray}  
which is the 2D wavefunction of two interacting bosons confined in a radial trap \cite{Bougas}. 
Here, $U(a,b,z)$ is the confluent hypergeometric function of the second kind \cite{Stegun}. 
Then, by plugging Eq. \eqref{2D_wavefunction} into Eq. \eqref{energy}, we retrieve the known 2D energy spectrum \cite{Bougas}
\begin{equation}
\psi\left(\frac{f(E)}{2}\right)=\ln \left(\frac{1}{2a_{\textrm{2D}}^2}\right)+2\ln 2-2\gamma,
\label{2D_spectrum}
\end{equation}
with $\psi(z)$ being the digamma function \cite{Stegun}. 

For convenience, in the following, we will refer to the states with energy less than the zero point energy, $E_0\equiv\frac{\alpha+1}{2}$, as bound states \cite{Calarco,Calarco2}. 
The energetically higher-lying states will be subsequently labeled the ground state, first excited state and so forth. 
Additionally, the energetic order of the eigenstates will be denoted by the subscripts 0 for the bound state, 1 for the ground state and in general $i$ denoting the $(i-1)$-th excited state. 
This labeling of the energies is explicitly showcased in Fig. \ref{Fig:Comparison} (a), and then it is omitted for brevity. Furthermore, a black dashed line is included to indicate the zero point energy. 

Figure \ref{Fig:Comparison} illustrates the two extreme regimes, namely the 2D case, for $\alpha=1$ [Fig. \ref{Fig:Comparison} (a)], and the quasi-1D 
case, for $\alpha \gg 1$ [Fig. \ref{Fig:Comparison} (b)]. 
In the quasi-1D regime, the spectrum of Eqs. \eqref{energy_spectrum2} and \eqref{num_continuation} is shown for $\alpha=10$ and compared with the energy spectrum directly derived from 
Eq. \eqref{spectrum_1D} for the 1D case. 
The two resulting energy spectra are presented together for a varying $g_{\textrm{1D}}$ in Fig. \ref{Fig:Comparison} (b). 
The zero point energy is put to $\frac{\alpha+1}{2}$. 
As it can be seen, regarding the excited states there is a perfect match for all values of $g_{\textrm{1D}}$. We should note however that for $\abs{g}>5$ there is a slight deviation between the two energies, which is of the order of $2\%$ at infinite $g_{\textrm{1D}}$. For a larger anisotropy, this discrepancy becomes smaller, for instance it is of the order of $0.5 \%$ at $\alpha=100$. 
Deviations between the two spectra arise also for the bound state in the attractive interaction regime, and in particular for large negative interactions  $g_{\textrm{1D}}<-10$ they become of the order of $15 \%$.
The aforementioned discrepancy, is due to the fact that the bound state in the pure 1D system exhibits a lower energy compared to the corresponding 2D setup.  
Indeed, the 2D system possesses bound states both in the attractive and the repulsive interaction regimes \cite{Bougas,Drummond} and for attractive couplings 
the energy of the bound state remains finite independently of the negative value of the interaction strength, see Fig. \ref{Fig:Comparison} (a). For positive values though the energy of the bound state is not bounded from below. 
This is in sharp contrast to the pure 1D system where the energy of the bound state in the attractive regime diverges at very strong 
interactions \cite{Laura}. 
As we shall discuss in the following, the energy gap between the bound and the ground states increases as the anisotropy parameter acquires larger values. 
However, for a larger value of $\alpha$ the above-mentioned discrepancy between the energies of the bound states in strictly 1D as compared to 2D (see Fig. \ref{Fig:Comparison} (b)) becomes smaller and occurs for stronger attractions. 
Note also that in Fig. \ref{Fig:Comparison} (b) there is a bound state in the repulsive interaction regime, having an energy much lower than the energy of the other states of the spectrum 
and is way below the shown energy scales.

\subsection{Energy dependence on the anisotropy parameter}

To expose the dependence of the eigenenergies on the anisotropy parameter $\alpha$, corresponding energy spectra are shown in Fig. \ref{Fig:Spectra} for different values of $\alpha$ thus accessing the 
dimensional crossover from the quasi-1D to the 2D regime. 
Evidently, in all cases the energy spacing among the different eigenstates is not equal, in contrast to the 2D case [Fig. \ref{Fig:Comparison} (a)], and greatly depends on $\alpha$. 
This behavior is anticipated by the expression of the energy for zero interactions, namely $E=2(n+\alpha m)+\frac{\alpha+1}{2}, \, n,m \, \in N$. 
For integer values of $\alpha$, the energy spacing between consecutive energy states becomes larger every $\alpha$-th state in both the attractive and the repulsive interaction regimes starting from the ground state, see for instance Figs. \ref{Fig:Spectra} (b) and (d). 
However, for non-integer $\alpha$ values, the energy spacings become more irregular as depicted in Figs. \ref{Fig:Spectra} (a), (c) and (e). 
For instance, at $\alpha=1.1$ and $g=0$ [Fig. \ref{Fig:Spectra} (a)], the energetical difference between the third and the fourth excited states is 
$2\alpha=0.2$. 
We should mention here that qualitatively similar results have been reported also for two bosons confined in a 3D anisotropic trap \cite{Calarco,Calarco2}. 
Moreover, the energy gap between the bound and the ground state increases for a larger anisotropy parameter independently of the sign of the interaction strength, see Figs. \ref{Fig:Spectra} (a)-(f). 

\begin{figure}[t!]
\centering
\includegraphics[width=0.47 \textwidth]{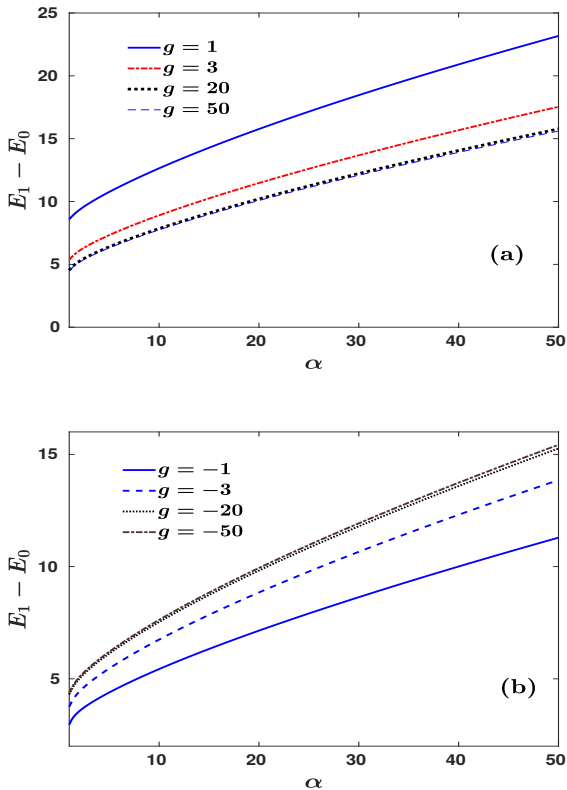}
\caption{Energy difference between the bound and the ground states, $E_1-E_0$, at different (a) repulsive and (b) attractive 2D interaction strengths (see legends) for varying anisotropy 
parameter $\alpha$. For all observables dimensionless units are adopted.}
\label{Fig:energy_diff}
\end{figure}

The energy of the bound states is shifted upwards for an increasing value of $\alpha$ due to the increase of the zero point energy, $\frac{\alpha+1}{2}$. 
To elaborate on the impact of the anisotropy parameter on the energy gaps we depict in Fig. \ref{Fig:energy_diff} the energy difference between the bound and the ground state, i.e. $E_1-E_0$, 
as a function of $\alpha$ for various repulsive [Fig. \ref{Fig:energy_diff} (a)] and attractive [Fig. \ref{Fig:energy_diff} (b)] interactions. 
We observe that the aforementioned energy difference increases for large $\alpha$ independently of the interactions and it does not saturate, e.g. at $\alpha=200$ and for $g=3$ $E_1-E_0=38.97$. 
Moreover, on the repulsive interaction regime [Fig. \ref{Fig:energy_diff} (a)], when $\alpha$ is kept constant, $E_1-E_0$ takes larger values at weak interactions. 
This is due to the divergence of the energy of the bound state close to the non-interacting limit of the repulsive interaction regime \cite{Zinner2,Drummond}. 
Also deep into the quasi-1D regime, i.e. $\alpha \gg 1$, the bound state is largely separated from the other states of the energy spectrum for all interaction strengths. 
On the attractive side [Fig. \ref{Fig:energy_diff} (b)], at fixed $\alpha$, the energy gap $E_1-E_0$ is larger at stronger attractions. 
For fixed attractive interaction $g$, $E_1-E_0$ becomes larger as the anisotropy parameter increases. 
Recall that for $g=0$ the energy of the bound state is always $\frac{\alpha+1}{2}$, i.e. it crosses the bound state threshold [see Fig. \ref{Fig:Comparison} (a)], and hence it is connected with $E_1$ at the repulsive side of the spectrum [Figs.  \ref{Fig:Spectra} (a)-(f)].

\begin{figure}[t!]
\centering
\includegraphics[width=0.47 \textwidth,keepaspectratio]{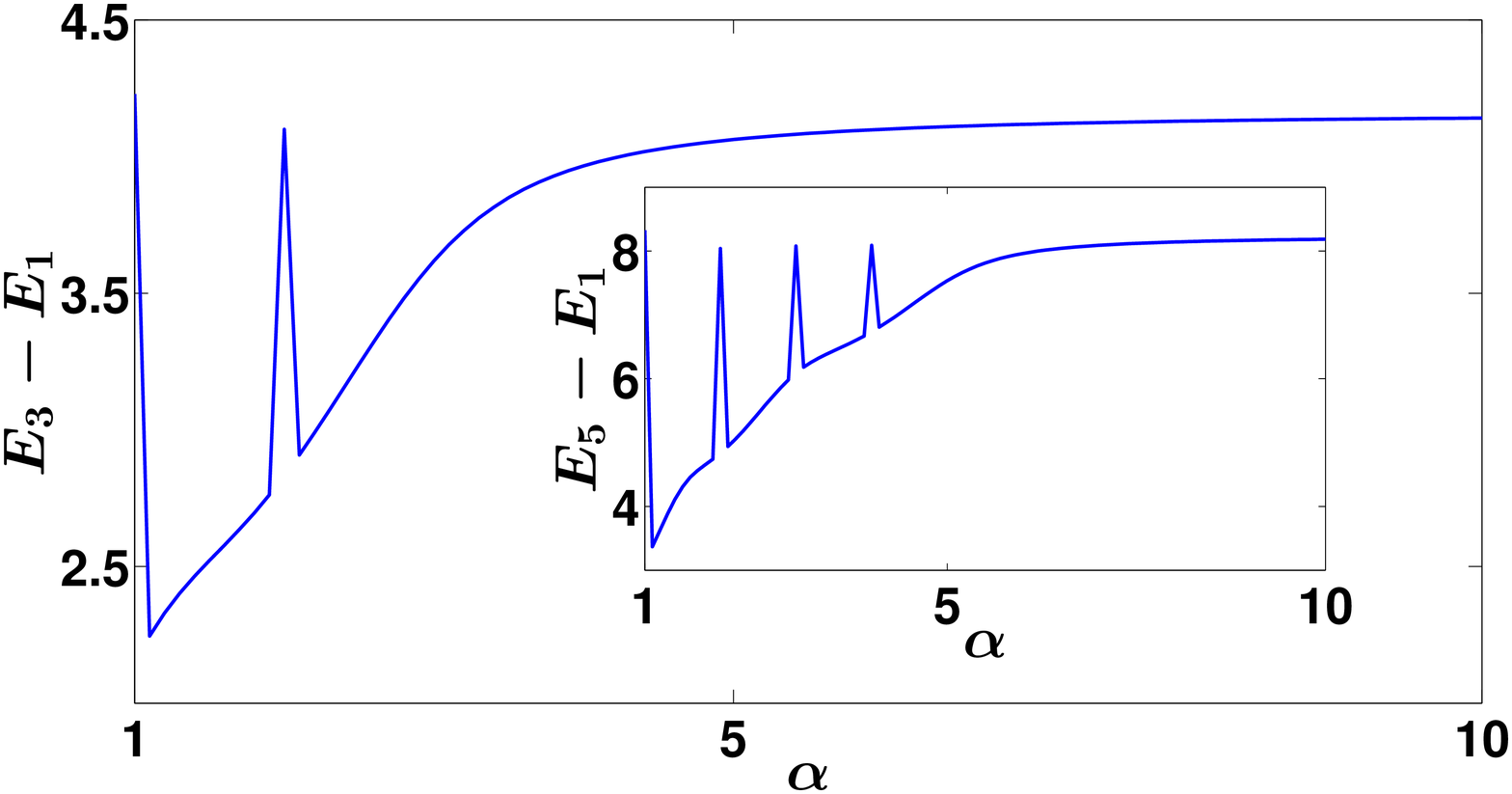}
\caption{Energy difference between the second excited and the ground state, namely $E_3-E_1$ for increasing anisotropy parameter $\alpha$. 
The inset presents the energy difference between the fourth excited and the ground state, i.e. $E_5-E_1$, with respect to $\alpha$. 
In both cases the 2D interaction strength of the two bosons is $g=3$. In all cases the quantities displayed are in dimensionless units.}
\label{Fig:energy_high}
\end{figure}

Figure \ref{Fig:energy_high} displays the energy difference between the second excited and the ground state, $E_3-E_1$, as well as between 
the fourth excited and the ground state $E_5-E_1$ in the corresponding inset, exemplarily for $g=3$. 
In both cases, for small $0<\alpha<5$ the energy spacings feature jumps and subsequently saturate for adequately large $\alpha>9$. 
These energy jumps occur for integer values of $\alpha$ and depend on the level of the excited state, for instance there are two jumps in the main Fig. \ref{Fig:energy_high} 
and four jumps in the inset. 
For values of $\alpha$ a little bit smaller or larger than these integer values, the energy gaps between the states decrease, see e.g. Fig. \ref{Fig:Spectra} (a) and hence the aforementioned jumps 
are manifested in the energy difference between excited states and the ground state. 
However, for anisotropies higher than the level of the examined excited state, the energy gap with the ground state saturates, because the change in the energy spacing occurs 
at even higher excited states. 
This is the case for the fourth excited state in Figs. \ref{Fig:Spectra} (d)-(f). 
We finally remark that for other interaction strengths of either sign, $E_3-E_1$ and $E_5-E_1$ exhibit a similar to the above-described behavior.

\section{Eigenstate analysis} \label{wavefunctions}

\subsection{Two-body wavefunction}

To acquire complete knowledge on the stationary properties of the system, we next determine the two boson wavefunction. 
The starting point is Eq. \eqref{wavefunction_integral}, where the integral is convergent for $f(E)>0$. 
However, it is advantageous to establish a more convenient form of $\Psi_{\textrm{rel}}(x,y)$ in order to span the entire energy spectrum. 
To this end, one can utilize the wavefunction ansatz introduced in Eq. \eqref{ansatz} along with the underlying expansion coefficients [Eq. \eqref{coeffs}]. 
Indeed, by expressing the denominator of Eq. \eqref{coeffs} in an integral representation, see Eq. \eqref{Int_rep}, and performing a single out of the two summations 
with the aid of the Mehler identity \cite{Ismail}, the two boson wavefunction of the relative coordinate takes the simplified form 
\begin{gather}
\Psi_{\textrm{rel}}(x,y)=\frac{B\sqrt{\alpha}}{\pi}e^{-(x^2+\alpha y^2)/2} \nonumber \\ 
\sum_{m=0}^{\infty} \frac{H_m(0)H_m(\sqrt{\alpha }y)\Gamma\left(\frac{\alpha m-\mathcal{E}}{2}\right)}{2^{m+1}m!} U\left(\frac{\alpha m-\mathcal{E}}{2},\frac{1}{2},x^2\right),
\label{Waves_simpl}
\end{gather}
where $\mathcal{E}=E-(\alpha+1)/2$. 
In practice, this summation is truncated when numerically calculating the wavefunction, with an upper bound which is chosen such that convergence is 
achieved \cite{Laura}. 
Note that the wavefunction in real space exhibits a logarithmic divergence close to the origin $x=y=0$, as already argued in Eq. \eqref{gamma_expansion}.  
However, the wavefunction of Eq. \eqref{Waves_simpl} cannot capture this behavior when truncating the infinite summation. 
Indeed, inserting $x=y=0$ in Eq. \eqref{Waves_simpl}, the wavefunction does not converge as we increase the cutoff in the summation. 
Moreover, the normalization constant $B$ can be easily determined analytically if we express the confluent hypergeometric function $U(a,b,x)$ in terms of parabolic cylinder 
functions $D_z(x)$ \cite{Stegun}.
For this choice, the integration can be performed analytically \cite{Gradshteyn} resulting in 
\begin{equation}
\begin{split}
B^{-2}=&\frac{\sqrt{\alpha}}{\sqrt{\pi}}\sum_{m=0}^{\infty}\frac{H_m(0)^2\Gamma\left(\frac{\alpha m-\mathcal{E}}{2}\right)}{2^{m+2}m!\Gamma\left(\frac{\alpha m-\mathcal{E}}{2}+\frac{1}{2}\right)}  \\
&  \times \left[\psi\left(\frac{1}{2}-\frac{\mathcal{E}-\alpha m}{2}\right)-\psi\left(-\frac{\mathcal{E}-\alpha m}{2}\right) \right],
\label{norma}
\end{split}
\end{equation} 
which corresponds to the analytical expression of the normalization coefficients.
\begin{figure*}
\centering
\includegraphics[width=0.7 \textwidth]{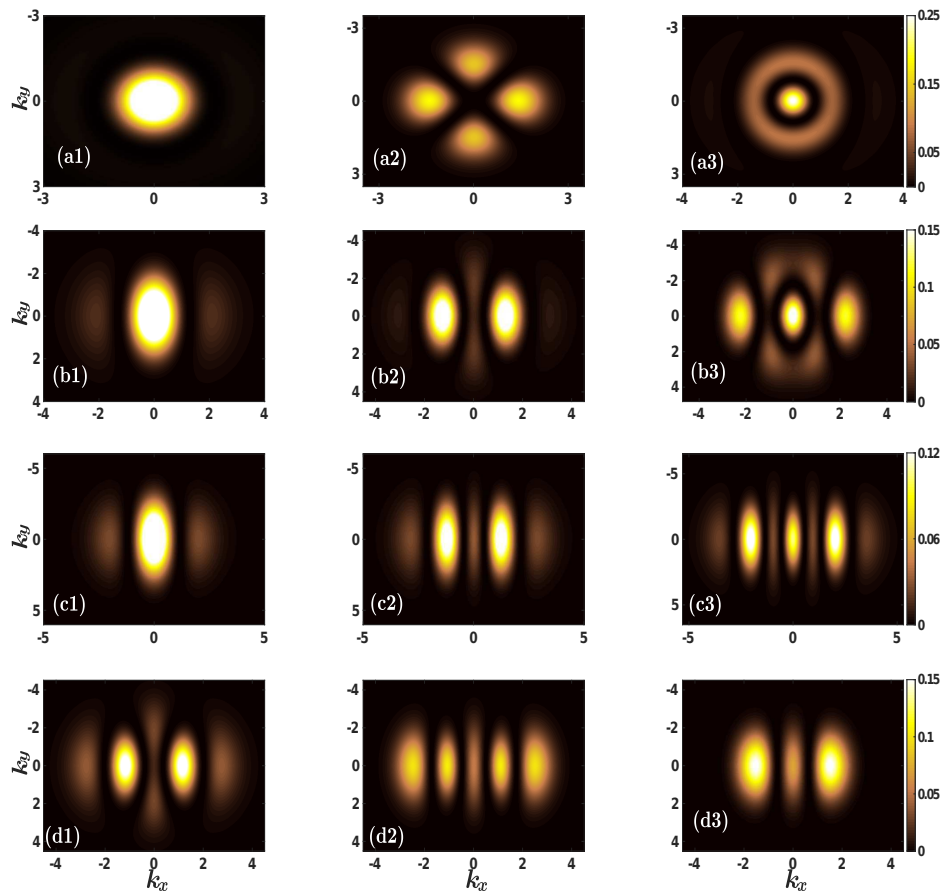}
\caption{Momentum distributions $\abs{\Psi_{\textrm{rel},j}(k_x,k_y)}^2$ for $\alpha=1.1$ ((a)-series), $\alpha=2.5$ ((b)-series) and $\alpha=5$ ((c)-series). The subindices (1,2,3) stand for the ground, first and second excited states respectively. 
All cases refer to interaction strength $g=1$. 
$\abs{\Psi_{\textrm{rel},2}(k_x,k_y)}^2$ of the first excited state ($j=2$) for $\alpha=2.5$ at (d1) $g=-1$, (d2) $g=-0.1$ and (d3) $g=0.1$. For all observables dimensionless units are adopted.}
\label{Fig:Momenta}
\end{figure*}

As pointed out in Section \ref{Energy Spectra}, the 2D wavefunction can be easily retrieved when $\alpha=1$, see Eq. \eqref{2D_wavefunction}. 
In the following, the wavefunction will be evaluated and further investigated deep into the quasi-1D regime i.e. in the case of $\alpha \gg 1$. 
Starting from Eq. \eqref{wavefunction_integral}, we note that in this case the wavefunction is elongated in the $x$ direction. 
Thus, in order to avoid the logarithmic divergence appearing at $x=y=0$ we shall restrict ourselves to $y=0$ and $x\gtrsim l_x$. 
With these simplifications Eq. \eqref{wavefunction_integral} is rewritten as
\begin{equation}
\begin{split}
\Psi_{\textrm{rel}}\left(x,0\right) &\approx \frac{B\sqrt{\alpha}}{2\pi} e^{-x^2/2}  \\&\int_{0}^{\infty} dw\, \textrm{exp}\left \{ -\frac{x^2e^{-w}}{1-e^{-w}} \right \}\frac{e^{-wf(E)/2}}{\sqrt{1-e^{-w}}}.
\end{split}
\end{equation}
Note that the square root involving the anisotropy parameter $\alpha$ in Eq. \eqref{wavefunction_integral} can be neglected, since for $w \gg \frac{1}{\alpha}$ 
the exponent $e^{-\alpha w}$ tends to zero. 
Also, for $w\ll \frac{1}{\alpha}$, the $\frac{1}{w}$ divergence in Eq. \eqref{integral_split} is counterbalanced by the factor $e^{-x^2/w}$, and the 
entire integrand vanishes. 
Employing a change of variables, $z=\frac{e^{-w}}{1-e^{-w}}$, it is easy to show that the wavefunction of two interacting bosons in a quasi-1D trap \cite{Laura} takes the approximate form 
\begin{equation}
\Psi_{\textrm{rel}}\left(x,0\right) \approx \frac{B\sqrt{\alpha}}{2\pi}e^{-x^2/2}\Gamma\left(\frac{f(E)}{2}\right)U\left(\frac{f(E)}{2},\frac{1}{2},x^2\right).  
\end{equation} 
The approximate nature of this expression stems from the fact that we have restricted ourselves to the spatial region $x\gtrsim l_x$. 

\subsection{Momentum Distribution}

Consequently, it is straightforward to calculate the wavefunction in momentum space. 
To this end, we utilize its expansion in terms of the Hermite polynomials introduced in Eq. \eqref{ansatz} as well as an identity 
regarding their Fourier transform \footnote{$\mathcal{F}\left \{ e^{-\alpha x^2/2}H_n(x\sqrt{\alpha}) \right \}= \frac{(-\iu)^n}{\sqrt{\alpha}}e^{-k^2/(2\alpha)}H_n\left(\frac{k}{\sqrt{\alpha}}\right)$, where $\mathcal{F} \{ g(x)  \} $ denotes the Fourier transform of a function $g(x)$.}.  
Therefore, the wavefunction $\Psi_{\textrm{rel}}(k_x,k_y)$ in momentum space reads 
\begin{equation} 
\begin{split}
\Psi_{\textrm{rel}}&(k_x,k_y)=\frac{B}{\pi} e^{-(k_x^2+k_y^2/\alpha)/2} \\
& \times \sum_{n,m}(-\iu)^{n+m}\frac{H_n(0)H_m(0)H_n(k_x)H_m(\frac{k_y}{\sqrt{\alpha}})}{2^{n+m}n!m!(n+\alpha m -\mathcal{E})}.
\label{Wave_mom}\
\end{split}
\end{equation} 

Since the wavefunction in real space exhibits a logarithmic divergence at the origin ($x=y=0$), it is better to analyze the structure of the two boson wavefunction 
in momentum space. 
Figure \ref{Fig:Momenta} illustrates the momentum distribution $\abs{\Psi_{\textrm{rel},j}(k_x,k_y)}^2$ for different anisotropy parameters $\alpha=1.1$ [Figs. \ref{Fig:Momenta} (a$j$)], 
$\alpha=2.5$ [Figs. \ref{Fig:Momenta} (b$j$)] and $\alpha=5$ [Figs. \ref{Fig:Momenta} (c$j$)], regarding the ground ($j=1$) and higher excited states ($j=2,3$) at $g=1$. 
Independently of the energetic order of the state we observe that as the anisotropy parameter increases the momentum distribution is elongated along the $k_y$ 
direction, see e.g. Figs. \ref{Fig:Momenta} (a1), (b1), (c1). 
This elongation occurs since the momentum distribution is more long-ranged for $k_y$ than $k_x$, according to the exponential decay given by Eq. \eqref{Wave_mom}. 
Additionally the momentum distribution for large anisotropies, see e.g. Figs. \ref{Fig:Momenta} (c1)-(c3), exhibits a multihump structure along the $k_x$ direction. 
This multihump structure becomes more pronounced for energetically higher excited states, compare for instance Figs. \ref{Fig:Momenta} (c2) and (c3). 
The latter behavior is attributed to the fact that the major contribution in the double summation of Eq. \eqref{Wave_mom} for high energies $\mathcal{E}$ (i.e. higher excited states), 
stems from higher order Hermite polynomials which are responsible for the observed multihump structure of the momentum distribution. 
Note also that for larger values of $\alpha$, a similar structure of the momentum distribution 
occurs as described in Figs. \ref{Fig:Momenta} (c1)-(c3) (not shown here for brevity). 
The momentum distribution of the first excited state ($j=2$) $\abs{\Psi_{\textrm{rel},2}(k_x,k_y)}^2$ for $\alpha=2.5$ is also presented at $g=-1$, $g=-0.1$ and $g=0.1$ in Figs. \ref{Fig:Momenta} (d1)-(d3). 
We deduce that as the attraction increases, $\abs{\Psi_{\textrm{rel},2}(k_x,k_y)}^2$ becomes more localized towards smaller values of $k_x$ while its outer humps are depleted, compare 
Figs. \ref{Fig:Momenta} (d1) and (d2). 
Also, in the vicinity of $g=0$ but on the attractive side, $\abs{\Psi_{\textrm{rel},2}(k_x,k_y)}^2$ develops an additional outer hump [Fig. \ref{Fig:Momenta} (d2)] 
compared to the momentum distribution for weak repulsions [Fig. \ref{Fig:Momenta} (d3)]. 
This is exactly due to the mismatch in the energy $\mathcal{E}_2$ in the vicinity of zero interactions, see Fig. \ref{Fig:Spectra} (c). 

A more complicated momentum structure of the first excited state ($j=2$) occurs for $\alpha=1.1$, where $\abs{\Psi_{\textrm{rel},2}(k_x,k_y)}^2$ displays a 
pedal-like structure [Fig. \ref{Fig:Momenta} (a2)]. 
We remark that for increasing anisotropy within the interval $\alpha \in [1.1,1.9]$, it is found that this pedal-like distribution becomes fainter along $k_y$ and 
more squeezed in the $k_x$ direction (not shown here). 
Moreover, these pedal patterns approach the origin i.e. $k_x=k_y=0$ for $\alpha\to1.9$. 
Let us also note that the energy of the first excited state at $\alpha=1.1$ and $g=1$ ($E=3.14633$) is close to the energy of a fermionic state with odd $n,m$ in the expression 
$E=n+\alpha m +\frac{\alpha+1}{2}$ ($E=3.15$). 
As $\alpha$ increases in the interval $\alpha \in [1.1,1.9]$, the energy of the first excited state at $g=1$ deviates significantly from the energy of the energetically 
closest fermionic state. 
The momentum distribution of the fermionic state exhibits also a pedal structure similar to the one presented in Fig. \ref{Fig:Momenta} (a2) but with a nodal line at $k_x=0$ and $k_y=0$. 
For $\alpha=1.9$, $\abs{\Psi_{\textrm{rel},2}(k_x,k_y)}^2$ shows a similar behavior to the one displayed in Fig. \ref{Fig:Momenta} (b2) for $\alpha=2.5$.  
At this value of $\alpha=2.5$, $\abs{\Psi_{\textrm{rel},3}(k_x,k_y)}^2$ of the second excited state ($j=3$) [Fig. \ref{Fig:Momenta} (b3)] exhibits populated tails for 
large $k_y$ values. 
As $\alpha$ increases these tails of the momentum distribution, in the $k_y$ direction, are suppressed and become apparent only for higher-lying excited states (not shown here for brevity).

\section{One-body densities} \label{one_body_stat} 

\begin{figure*}
\centering
\includegraphics[width= \textwidth]{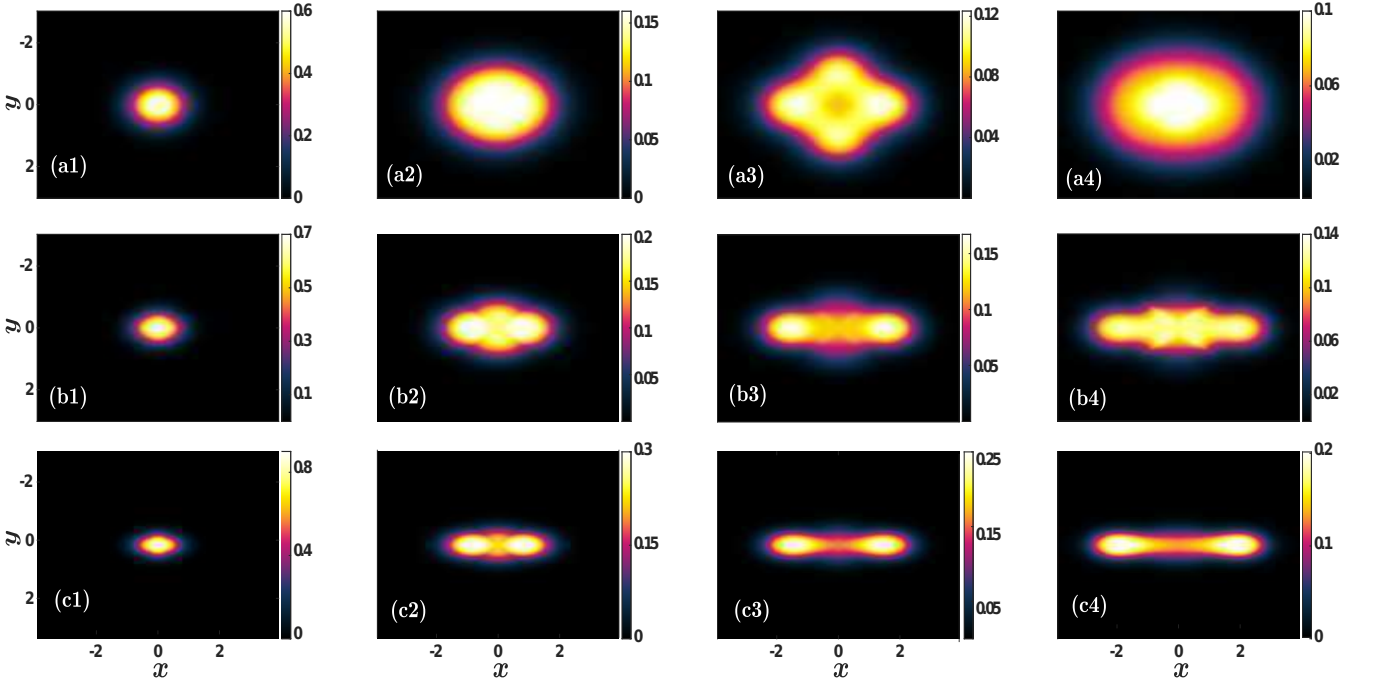}
\caption{One-body densities, $\rho^{(1)}(x_1,y_1)$ for $\alpha=1.1$ ((a)-series), $\alpha=2.5$ ((b)-series) and $\alpha=5$ ((c)-series). The subscripts (1,2,3,4) refer to the bound, ground, first and second excited states respectively. 
In all cases the interparticle interaction strength is $g=1$. All quantities shown are in dimensionless units.}
\label{Fig:Densities}
\end{figure*} 

Having at hand the two boson wavefunction for an arbitrary anisotropy parameter enables us to access all the properties of the system. 
As a case example, we shall investigate the corresponding one-body densities $\rho^{(1)}(x_1,y_1)$ for several states and anisotropies. 
The one-body density of two bosons reads \cite{Sakmann} 
\begin{eqnarray}
\rho^{(1)}(x_1,y_1)&=& \int dx_2 dy_2 |\Psi_{\textrm{c.m.}}\left(X(x_1,x_2),Y(y_1,y_2)\right) \nonumber \\
& & \times \Psi_{\textrm{rel}}\left(x(x_1,x_2),y(y_1,y_2)\right)|^2 \label{density_prel}
\end{eqnarray} 
For the relative coordinate wavefunction, we employ the expansion of Eq. \eqref{Waves_simpl}. 
The center-of-mass wavefunction resides in its ground state, as was discussed in Section \ref{Setup}. 
To perform the integral appearing in Eq. (\ref{density_prel}) we utilize the coordinate transformations of the center-of-mass and relative coordinates, and therefore express all variables 
in terms of the positions of the two bosons. 
In this way, the one-body density reads
\begin{widetext} 
\begin{gather}
\rho^{(1)}(x_1,y_1)=\frac{B^2\alpha^{3/2}}{\pi^3}e^{-(x_1^2+\alpha y_1^2)}\sum_{n,m}f(n)f(m)\overbrace{\int_{-\infty}^{+\infty}dy_2\, e^{-\alpha y_2^2}\, H_n\left(\sqrt{\alpha}\frac{y_1-y_2}{\sqrt{2}}\right)H_m\left(\sqrt{\alpha} \frac{y_1-y_2}{\sqrt{2}}\right)}^J \nonumber \\
\times \int_{-\infty}^{+\infty} dx_2\, e^{-x_2^2}\,U\left(\frac{\alpha m-\mathcal{E}}{2},\frac{1}{2},\frac{(x_1-x_2)^2}{2}\right)U\left(\frac{\alpha n-\mathcal{E}}{2},\frac{1}{2},\frac{(x_1-x_2)^2}{2}\right),
\end{gather}
\end{widetext}
with $f(n)=\frac{H_n(0)\Gamma\left(\frac{\alpha n-\mathcal{E}}{2} \right)}{2^{n+1}\Gamma(n+1)}$. 
The first integral denoted by $J$ can be calculated analytically by using the transformation $y_2 \rightarrow \sqrt{\alpha}y_2$, 
and subsequently the substitution $y_2=y_1\sqrt{\alpha}-y_2$ \cite{Gradshteyn}. 
Then, the integral is 
\begin{equation}
J=\frac{\sqrt{\pi}}{\sqrt{\alpha}}\sum_{k=0}^{\textrm{min}(n,m)}2^kk!\binom{m}{k}\binom{n}{k}\frac{1}{2}^{\frac{m+n}{2}-k}H_{m+n-2k}(y_1\sqrt{\alpha}).
\end{equation} 

Figure \ref{Fig:Densities} illustrates the one-body densities of the bound, ground, first and second excited states at $g=1$ when $\alpha=1.1$ [Fig. \ref{Fig:Densities} (a1)-(a4)], 
$\alpha=2.5$ [Fig. \ref{Fig:Densities} (b1)-(b4)] and $\alpha=5$ [Fig. \ref{Fig:Densities} (c1)-(c4)]. 
If $\alpha\approx1$, $\rho^{(1)}(x,y)$ of the higher-lying excited states [Fig. \ref{Fig:Densities} (a2)-(a4)] tends to show an almost isotropic distribution along the $x$ and $y$ directions. 
On the other hand, for a large anisotropy parameter $\alpha$ the 1D limit is approached and therefore $\rho^{(1)}(x,y)$ becomes more elongated in the $x$ direction [Figs. \ref{Fig:Densities} (c1)-(c4)]. 
Indeed, as the anisotropy $\alpha$ increases, the one-body densities of the ground and higher excited states develop a prominent two-hump structure in the elongated $x$ direction, see for instance 
Figs. \ref{Fig:Densities} (c2)-(c4) where $\alpha=5$. 
This is reminiscent of the behavior of the one-body densities of two bosons confined in a 1D harmonic trap \cite{Laura,Bolsinger}. 
Entering the intermediate anisotropy regime, e.g. $\alpha=2.5$ [Fig. \ref{Fig:Densities} (b1)-(b4))], $\rho^{(1)}(x,y)$ exhibits population tails along the $y$ direction as well. 
The two hump structure of $\rho^{(1)}(x,y)$ is present in the ground [Fig. \ref{Fig:Densities} (b2)] and the first excited state [Fig. \ref{Fig:Densities} (b3)], but disappears 
in the second excited state [Fig. \ref{Fig:Densities} (b4)] and in higher excited states as well (not shown). 
However, for small anisotropies [Figs. \ref{Fig:Densities} (a1)-(a4), $\alpha=1.1$], the one-body density resembles the structure of the corresponding pure 2D case \cite{Bougas}. 
The only exception is the first excited state [Fig. \ref{Fig:Densities} (a3)] which features a small density dip at the center $x=y=0$. 
Recall that this latter state corresponds to the pedal-like structure of the momentum distribution depicted in Fig. \ref{Fig:Momenta} (a2). 
Finally, the one-body density of the bound states [Figs. \ref{Fig:Densities} (a1), (b1) and (c1)], is more elongated in the $x$ direction and somewhat localized near the origin, $x=y=0$. 
The latter is due to the fact that the bound state is strong in the repulsive interaction regime, as was discussed in Section \ref{Energy Spectra} [see Fig. \ref{Fig:Spectra}].

\section{Tan Contacts} \label{Tan_stat}

In Section \ref{Energy Spectra}, it was argued that at interparticle distances much smaller than $l_x$, the two boson wavefunction develops a logarithmic divergence. 
This behavior is caused by the contact interaction in 2D, see also Eq. \eqref{Pot}, which can also be expressed as a boundary condition for the wavefunction at zero interparticle 
distances \cite{Pricoup,Combescot}, where the Tan contact is defined \cite{Tan1,Tan2,Tan3,Werner,Vanja,Jin1,Jin2}.
In this section we measure the Tan contact as a function of the anisotropy parameter $\alpha$ for various eigenstates and several interaction strengths. 

The Tan contact, $\mathcal{D}$, is defined from the momentum distribution in the limit of very large momenta, namely $\abs{\Psi(k)}^2\stackrel{k\rightarrow \infty}{\longrightarrow} \frac{\mathcal{D}}{k^4}$, 
in all dimensions \cite{Werner,Barth,Vignolo}. 
Since the wavefunction at small interparticle distances depends only on the radius $r^2=x^2+y^2$ [see also Eq. \eqref{gamma_expansion}], and the Tan contact is determined by the behavior 
of the wavefunction at $r\to0$ \cite{Corson}, $\mathcal{D}$ is isotropic, i.e. it does not depend on the $x$ or $y$ direction. The contact reads [for details see Appendix \ref{Ap:Tan}]
\begin{equation}
\mathcal{D}(\alpha,\mathcal{E})=\frac{B^2(\alpha,\mathcal{E})}{4\pi^4}.
\label{stat_contact}
\end{equation}
Therefore, this Tan contact is essentially defined by the normalization constant $B(\alpha,\mathcal{E})$ of the wavefunction [Eq. \eqref{norma}] and refers to the two-body state which is in turn 
characterized by the anisotropy parameter $\alpha$ and the energy $\mathcal{E}$. 
In the quasi-1D limit, i.e. $\alpha \gg 1$, we obtain the following relation [for details see Appendix \ref{Ap:Tan}]
\begin{equation}
\mathcal{D_{\textrm{2D}}}=l_y \sqrt{\pi}\mathcal{D}_{\textrm{1D}}.
\label{Rel_cont} 
\end{equation} 
As a consequence the 2D and the 1D contacts are linked via a geometric factor $\sqrt{\pi}$ and the harmonic oscillator length of the strongly confined direction. 
Note that the three-dimensional contact is also related to the lower dimensional ones through specific geometric factors and the oscillator lengths in the tightly confined 
directions \cite{Valiente,He,Decamp}. 
In what follows, we shall explore $\mathcal{D}(\alpha,\mathcal{E})$ rescaled by the factor $1/l_y$ (or $\sqrt{\alpha}$ in harmonic oscillator units) in order to expose the 
connection between the contacts in 1D and 2D, and subsequently showcase the saturation of the $\mathcal{D}_{2D}$ for large values of $\alpha$ towards the value of the 1D contact. 

\begin{figure}[t!]
\centering
\includegraphics[width=0.47 \textwidth]{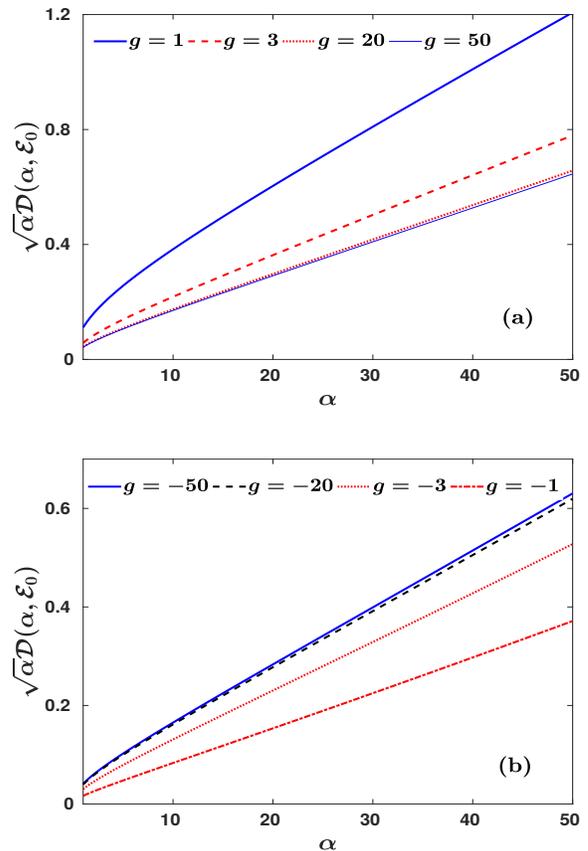}
\caption{Rescaled Tan contact $\sqrt{\alpha}\mathcal{D}(\alpha,\mathcal{E}_0)$ of the bound state at various (a) repulsive and (b) attractive interaction strengths (see legends) for increasing 
anisotropy parameter $\alpha$. In all cases the quantities displayed are in dimensionless units.}
\label{Fig:Contacts_bound}
\end{figure}

Figure \ref{Fig:Contacts_bound}, depicts $\sqrt{\alpha}\mathcal{D}(\alpha,\mathcal{E}_0)$ of the bound states with respect to $\alpha$, for both repulsive [Fig. \ref{Fig:Contacts_bound} (a)] 
and attractive [Fig. \ref{Fig:Contacts_bound} (b)] interaction strengths. 
We observe that for increasing $\alpha$ and independently of the interaction strength, the contact takes larger values and does not saturate.
This enhancement of two-body short-range correlations is attributed to the fact that the bound states in the repulsive as well as in the attractive regime become more deeply bound as the anisotropy inreases, see also Fig. \ref{Fig:Spectra}. 
Furthermore, at fixed anisotropy $\alpha$ and weak interparticle interactions [Fig. \ref{Fig:Contacts_bound} (a)], the contact is enhanced compared to the one for larger interaction strengths. 
This can be explained from the fact that the bound state diverges for weak repulsive interactions [see Figs. \ref{Fig:Spectra} (a)-(f)] and therefore the degree of short-range correlations is enhanced. 
On the contrary, for attractive interactions [Fig. \ref{Fig:Contacts_bound} (b)], the contact increases as the interactions become more attractive, while $\alpha$ is kept fixed. 
Indeed, inspecting Figs. \ref{Fig:Spectra} (a)-(f) reveals that for a stronger attraction the contribution of the bound state becomes substantial.
\begin{figure}[t!]
\centering
\includegraphics[width=0.47 \textwidth]{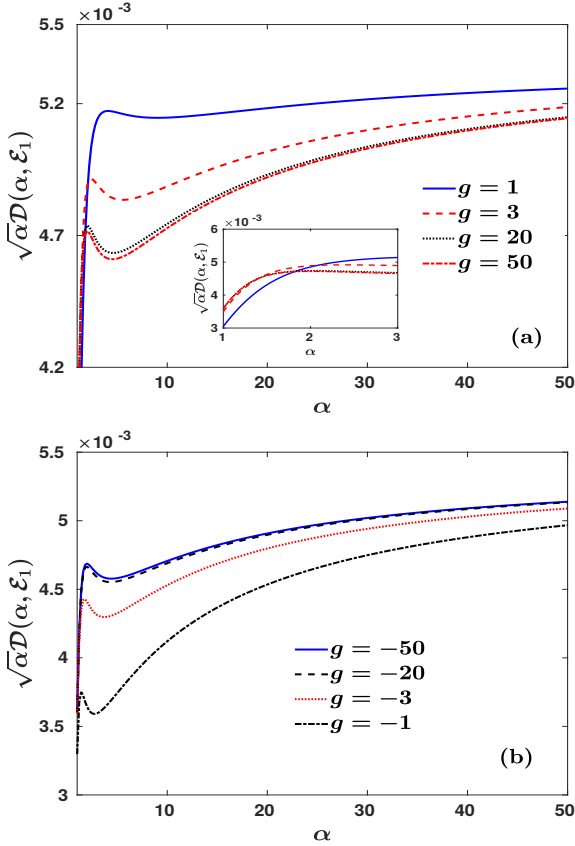}
\caption{Rescaled Tan contact $\sqrt{\alpha}\mathcal{D}(\alpha,\mathcal{E}_1)$ of the ground states at different (a) repulsive and (b) attractive interactions (see legends) for varying 
anisotropy $\alpha$. 
The inset in (a) presents a magnification of $\sqrt{\alpha}\mathcal{D}(\alpha,\mathcal{E}_1)$ within the anisotropy interval $\alpha \in [1,3]$. For all observables dimensionless units are adopted.}
\label{Fig:Contacts_ground}
\end{figure}

The rescaled contact $\sqrt{\alpha}\mathcal{D}(\alpha,\mathcal{E}_1)$ of the ground state as a function of the anisotropy parameter $\alpha$ is illustrated in Fig. \ref{Fig:Contacts_ground} for 
various repulsive [Fig. \ref{Fig:Contacts_ground} (a)] and attractive [Fig. \ref{Fig:Contacts_ground} (b)] interactions. 
As it can be seen, in contrast to Fig. \ref{Fig:Contacts_bound}, $\sqrt{\alpha}\mathcal{D}(\alpha,\mathcal{E}_1)$ features an initial growth and then it saturates to a value that is proportional 
to the 1D contact [Eq. \eqref{Rel_cont}] for all coupling strengths. 
Initially at $\alpha=1$, the contact possesses a higher value for strong repulsions \cite{Jin2}, see Fig. \ref{Fig:Contacts_ground} (a). 
However, this behavior is reversed as the anisotropy increases, and $\sqrt{\alpha}\mathcal{D}(\alpha,\mathcal{E}_1)$ acquires larger values for weaker repulsive interactions, compare for instance $g=1$ and $g=3$ 
in Fig. \ref{Fig:Contacts_ground} (a) for $\alpha \geq 5$. 
This latter feature is better visualized in the inset of Fig. \ref{Fig:Contacts_ground} (a) where $\sqrt{\alpha}\mathcal{D}(\alpha,\mathcal{E}_1)$ is showcased within the anisotropy interval 
$\alpha \in [1,3]$ and the aforementioned inverted behavior occurs at $\alpha \simeq 2$. 
Indeed, for increasing $\alpha$ we enter deep into the quasi-1D regime and therefore one should use the corresponding 1D interaction strength being related to its 2D counterpart via Eq. \eqref{couplings}. 
This relation maps the repulsive 2D interactions to attractive 1D interactions for large values of the anisotropy $\alpha$. 
For instance, Eq. \eqref{couplings} provides the mapping $g_{\textrm{2D}}=(1,3,20,50)\mapsto g_{\textrm{1D}}=(-6.403,-5.045,-4.628,-4.588)$ for $\alpha=10$. 
Similarly, for attractive interactions an increasing behavior of the short-range two-body correlations as captured by $\sqrt{\alpha}\mathcal{D}(\alpha,\mathcal{E}_1)$ occurs and then a tendency of saturation is 
observed independently of the coupling strength [Fig. \ref{Fig:Contacts_ground} (b)]. 
When $\alpha$ is fixed, $\sqrt{\alpha}\mathcal{D}(\alpha,\mathcal{E}_1)$ acquires larger values for a stronger attraction. 
Here, Eq. \eqref{couplings} maps the strong 2D attraction to the strong 1D attraction, for large anisotropies. 
Explicitly this mapping reads $g_{\textrm{2D}}=(-50,-20,-3,-1) \mapsto g_{\textrm{1D}}=(-4.535,-4.496,-4.162,-3.542)$ for $\alpha=10$.

Another interesting observation is that $\sqrt{\alpha}\mathcal{D}(\alpha,\mathcal{E}_1)$ shows a peak within $\alpha \in [2,4]$, see Figs. \ref{Fig:Contacts_ground} (a) and (b). 
Indeed, for a small anisotropy parameter the energy of the ground state, $\mathcal{E}_1$, increases in both the repulsive and the attractive interaction regimes for larger $\alpha$ satisfying $\alpha \in [2,4]$. 
Hence, the Tan contact is also enhanced in this $\alpha$ interval. 
Note also that $\sqrt{\alpha}\mathcal{D}(\alpha,\mathcal{E}_1)$ for fixed $\alpha$ becomes smaller [larger] for increasing repulsive [attractive] 2D coupling strength, see Figs. \ref{Fig:Contacts_ground} (a) and (b). 
However, if $\alpha$ exceeds a critical value depending on $g$, we approach the quasi-1D region and Eq. \eqref{couplings} maps the 2D to the 1D coupling strength.  
In particular, for $\alpha \in [2,6]$, the 1D coupling becomes less attractive acquiring larger negative values for increasing $\alpha$. 
Hence, qualitatively $\sqrt{\alpha}\mathcal{D}(\alpha,\mathcal{E}_1)$ initially increases up to a point where the crossover to 1D starts to become important and then it decreases similarly to the absolute value of $g_{\textrm{1D}}$ \cite{Barth}. 
Subsequently, the 1D attraction is enhanced and the contact increases up to its saturation value.

\section{Interaction quench dynamics} \label{dynamics}

\subsection{Time evolution of the wavefunction}

Having analyzed the stationary properties of the two-boson system in the dimensional crossover from 2D to 1D we next proceed by investigating the 
resulting interaction quench dynamics of this setup for a fixed anisotropy parameter $\alpha$ and different postquench 2D interaction strengths $g$. 
As already discussed in Sec. \ref{Setup}, the center-of-mass wavefunction $\Psi_{\textrm{c.m.}}(X,Y)$ [Eq. (\ref{Wave_CM})] lies in the ground state and thus it is not affected by the interaction quench. 
Therefore, the center-of-mass wavefunction does not play any role in the description of the interaction quench dynamics and it will not be considered in the following analysis. 

To be more precise, in order to study the dynamics, the system is initially prepared in an eigenstate $\ket{\Psi_{\textrm{rel},i}^{\textrm{in}}(x,y;0)}$ at an initial interaction 
strength $g^{\textrm{in}}$ with energy $\mathcal{E}_i^{\textrm{in}}$ and at $t=0$ this coupling strength is suddenly changed (quenched) to a final (postquench) value $g$. 
Then, the time-evolution of the initial wavefunction reads
\begin{gather}
\ket{\Psi_{\textrm{rel},i}^{\textrm{in}}(x,y;t)}=e^{-\iu \hat{\mathcal{H}} t} \ket{\Psi_{\textrm{rel},i}^{\textrm{in}}(x,y;0)} \nonumber \\
=\sum_j e^{-\iu \mathcal{E}_j t} \ket{\Psi_{\textrm{rel},j}^f(x,y)}\underbrace{\braket{\Psi_{\textrm{rel},j}^f(x,y)| \Psi_{\textrm{rel},i}^{\textrm{in}}(x,y;0)}}_{d_{i,j}},
\label{T-wvaefunction}
\end{gather}
where the summation is performed over the eigenstates of the postquench Hamiltonian $\ket{\Psi_{\textrm{rel},j}^f(x,y)}$ with energy $\mathcal{E}_j$. 
The underlying overlap coefficients, $d_{i,j}$, are determined by employing the ansatz introduced in Eq. \eqref{ansatz} as well as the orthonormality of the non-interacting 
wavefunctions $\phi_n(x)$ and have the form
\begin{gather}
d_{i,j}= \frac{B_iB_j}{\mathcal{E}_i^{\textrm{in}}-\mathcal{E}_j}\sqrt{\frac{\alpha}{\pi}}\sum_{m \geq 0} \frac{H_m^2(0)}{2^{m+1}m!} \times \nonumber \\ 
\times \left[ \frac{\Gamma\left(\frac{\alpha m-\mathcal{E}_i^{\textrm{in}}}{2}\right)}{\Gamma\left(\frac{1+\alpha m -\mathcal{E}_i^{\textrm{in}}}{2}\right)} -\frac{\Gamma\left(\frac{\alpha 
m-\mathcal{E}_j}{2}\right)}{\Gamma\left(\frac{1+\alpha m-\mathcal{E}_j}{2}\right)} \right].\label{overlap_coef}
\end{gather} 
These overlap coefficients between the initial wavefunction, $\Psi_{\textrm{rel},i}^{\textrm{in}}(x,y;0)$, and a final eigenstate, $\Psi_{\textrm{rel},j}^f(x,y)$, determine the degree of 
participation of this postquench eigenstate in the dynamics.   

\subsection{Dynamical response of the system}

A well-known observable of interest that enables us to identify the dynamical response of the system to its external perturbation, herein an interaction quench, is the fidelity. 
The latter is defined by the overlap between the time-evolved and the initial wavefunction \cite{Gorin,Fogarty,Thies,Jannis,Goold}, namely 
\begin{equation}
F(t)=  \braket{\Psi_{\textrm{rel},i}^{\textrm{in}}| e^{-\iu \hat{\mathcal{H}}t}| \Psi_{\textrm{rel},i}^{\textrm{in}}}= \sum_j e^{-\iu \mathcal{E}_j t} \abs{d_{i,j}}^2.
\label{Fid}
\end{equation}
Evidently, $F(t)$ is tailored to estimate the instantaneous deviation of the system from its initial state. 
Below, in order to capture the mean dynamical response of the system after a quench we invoke the time-averaged fidelity i.e. $\abs{\bar{F}}= \lim_{T \rightarrow \infty}\frac{1}{T}\int_{0}^{T} dt \abs{F(t)}$.

\begin{figure}[t!]
\centering
\includegraphics[width= 0.47 \textwidth]{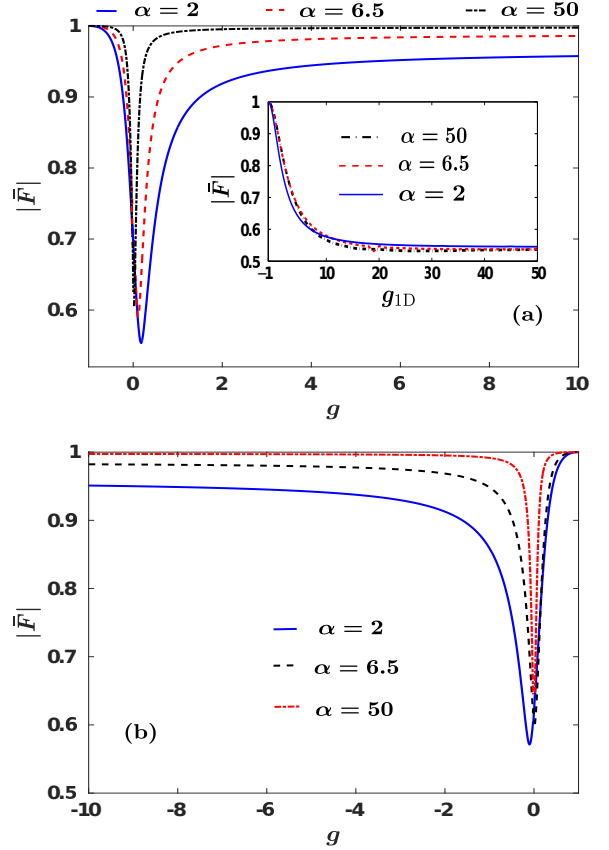}
\caption{Time-averaged fidelity $\abs{\bar{F}}$ as a function of the 2D interaction strength $g$ for various anisotropies (see legends). 
(a) The dynamics is triggered by following an interaction quench from the ground state of the system with $g^{\textrm{in}}=-1$ to larger interactions. (Inset) $|\bar{F}|$ following a quench from $g_{\textrm{1D}}^{\textrm{in}}=-1$ to larger 1D interactions for different anisotropies (see legend). 
(b) The quench is applied from the ground state of the two bosons with $g^{\textrm{in}}=1$ to smaller values of the interaction strengths. All quantities shown are in dimensionless units.}
\label{Fig:Mean_fid}
\end{figure}

The resulting $\abs{\bar{F}}$ following an interaction quench from the ground state either at $g^{\textrm{in}}=-1$ to repulsive postquench interactions is shown in Fig. \ref{Fig:Mean_fid} (a) 
or at $g^{\textrm{in}}=1$ towards the attractive regime is depicted in Fig. \ref{Fig:Mean_fid} (b) for various anisotropies namely $\alpha=2,\, 6.5$ and 50. 
In both quench scenarios and for all displayed anisotropies, $\abs{\bar{F}}$ drops from unity by developing a characteristic dip in the vicinity of zero postquench interactions indicating that the 
system is significantly perturbed for these values of $g$.
However, $\abs{\bar{F}}$ tends to approach values close to unity for large attractive or repulsive postquench interaction strengths $g$, evincing that the system remains 
close to its initial state. 
The above-described behavior of $\abs{\bar{F}}$ indicates the fact that the time-evolved two-body state in the vicinity of zero interactions is a non-trivial superposition containing many postquench eigenstates. 
However, for quenches to strong attractive or repulsive interactions the system populates a much smaller amount of postquench eigenstates and thus deviates from the initial state to a lesser extent compared to the $g=0$ case. 
For instance, the initial state $E_1$ at $g^{\textrm{in}}=-1$ is energetically close to the postquench $E_1$ at $g>1$ and therefore this eigenstate predominantly contributes to the time-evolved wavefunction. 
This is in contrast e.g. to the case of a quench to $g=0$ where both the $E_0$ and $E_1$ postquench eigenstates are energetically close to the initial $E_1$. 
The explicit contribution of the postquench eigenstates will be discussed below in detail. 

The width of the aforementioned dip of $\abs{\bar{F}}$ becomes more narrow as $\alpha$ increases and its location is displaced towards zero postquench interactions. 
Also, the minimum value of $\abs{\bar{F}}$ in the region of the dip increases for a larger anisotropy. 
Interestingly, for large postquench attractive or repulsive interactions, e.g. $\abs{g}=8$ in Figs. \ref{Fig:Mean_fid} (a) and (b), the system deviates more from its initial configuration 
as the anisotropy $\alpha$ becomes smaller. 
Furthermore, in both quench scenarios, as $\alpha$ increases, $\abs{\bar{F}}$ tends to saturate close to unity for smaller interaction strengths $g$, see Figs. \ref{Fig:Mean_fid} (a) and (b). 
This latter behavior stems from the underlying energy spectrum presented in Fig. \ref{Fig:Spectra} and the associated energy gaps. 
Indeed, as the anisotropy increases, the saturation of the energies to their values at $g=0$ occurs at smaller attractive or repulsive interactions. 
Therefore, by decreasing the anisotropy of the 2D system we can drive it out-of-equilibrium in a more efficient manner. 

To further expose the interplay between the 2D and the 1D effective coupling constants, we showcase in the inset of Fig. \ref{Fig:Mean_fid} (a) the dependence of $\abs{\bar{F}}$ on $g_{\textrm{1D}}$ for distinct values of $\alpha$. 
Here, the explicit relation between the $g_{\textrm{2D}}$ and the $g_{\textrm{1D}}$ [Eq. \eqref{couplings}] has been used. 
As before, initially, $g^{\textrm{in}}_{\textrm{1D}}=-1$ and the interaction quench is performed towards the repulsive regime. 
In all cases, i.e. independently of $\alpha$, $\abs{\bar{F}}$ exhibits a decreasing tendency for increasing $g_{\textrm{1D}}$ until it approaches a fixed value for large $g_{\textrm{1D}}$.    
Recall that the energy spacings among the involved eigenstates in 1D [Fig. \ref{Fig:Comparison} (b)] saturate only for very large attractive or repulsive interactions and thus $|\bar{F}|$ tends to a constant value after $g_{\textrm{1D}}>20$. 
For these values of $g_{\textrm{1D}}$ we approach the strongly interacting regime and the (time-averaged) overlap of the time-dependent two-body state with the initial one is very small \cite{Laura}. 
It is also worth mentioning that the deviation of $|\bar{F}|$ between $\alpha=6.5$ and $\alpha=50$ is very small. 
However for $\alpha=2$, where the quasi-1D limit is not well-established, $|\bar{F}|$ differs noticeably e.g. from the case of $\alpha=6.5$. 
Note again that the quasi-1D limit is adequately approached for $\alpha>10$, see also Fig. \ref{Fig:Comparison} (b). 
Therefore the involved energy spacings which are considerably different between $\alpha=2$ [Fig. \ref{Fig:Spectra} (b)] and $\alpha=6.5$ [Fig. \ref{Fig:Spectra} (e)], $\alpha=50$ result in the observed discrepancy of $\abs{\bar{F}}$ between the aforementioned values of $\alpha$. 

\subsection{Dynamics of the position variance along each spatial direction}

Due to the considered anisotropy of the 2D harmonic trap, different frequencies will be excited along the two spatial directions after the quench, thus yielding a much richer dynamics compared to the purely 
isotropic case, as has been reported in several experiments with anisotropic 3D traps \cite{Kottke}. 
To study the excitations in the different spatial directions of the trap, we resort to the frequency spectra of the spatial extent of the relative 
wavefunction along the $x$ and $y$ directions \cite{Simos1,Simos2,Thies}. 
The instantaneous spatial extent of the two-boson cloud in each spatial direction is given by the respective variances  
\begin{gather}
\braket{x^2(t)}= \int_{-\infty}^{\infty} dx dy \, x^2 \abs{\Psi_{\textrm{rel},i}^{\textrm{in}}(x,y;t)}^2  \\
\braket{y^2(t)}= \int_{-\infty}^{\infty} dxdy \, y^2 \abs{\Psi_{\textrm{rel},i}^{\textrm{in}}(x,y;t)}^2.
\label{mean_X} 
\end{gather}
These observables allow us to monitor the expansion and contraction of the bosonic cloud in the course of the time-evolution and also to identify the frequencies of the 
participating modes in the dynamics along each spatial direction. 
This can be achieved by utilizing the frequency spectra of $\braket{x^2(t)}$ and $\braket{y^2(t)}$, namely $F(\omega_{x})=\frac{1}{\sqrt{2\pi}}\int_{-\infty}^{\infty} dt\, e^{\iu \omega_{x} t} \braket{x^2(t)}$ 
and $F(\omega_{y})=\frac{1}{\sqrt{2\pi}}\int_{-\infty}^{\infty} dt\, e^{\iu \omega_{y} t} \braket{y^2(t)}$ respectively.

\begin{figure}[t!]
\centering
\includegraphics[width= 0.47 \textwidth]{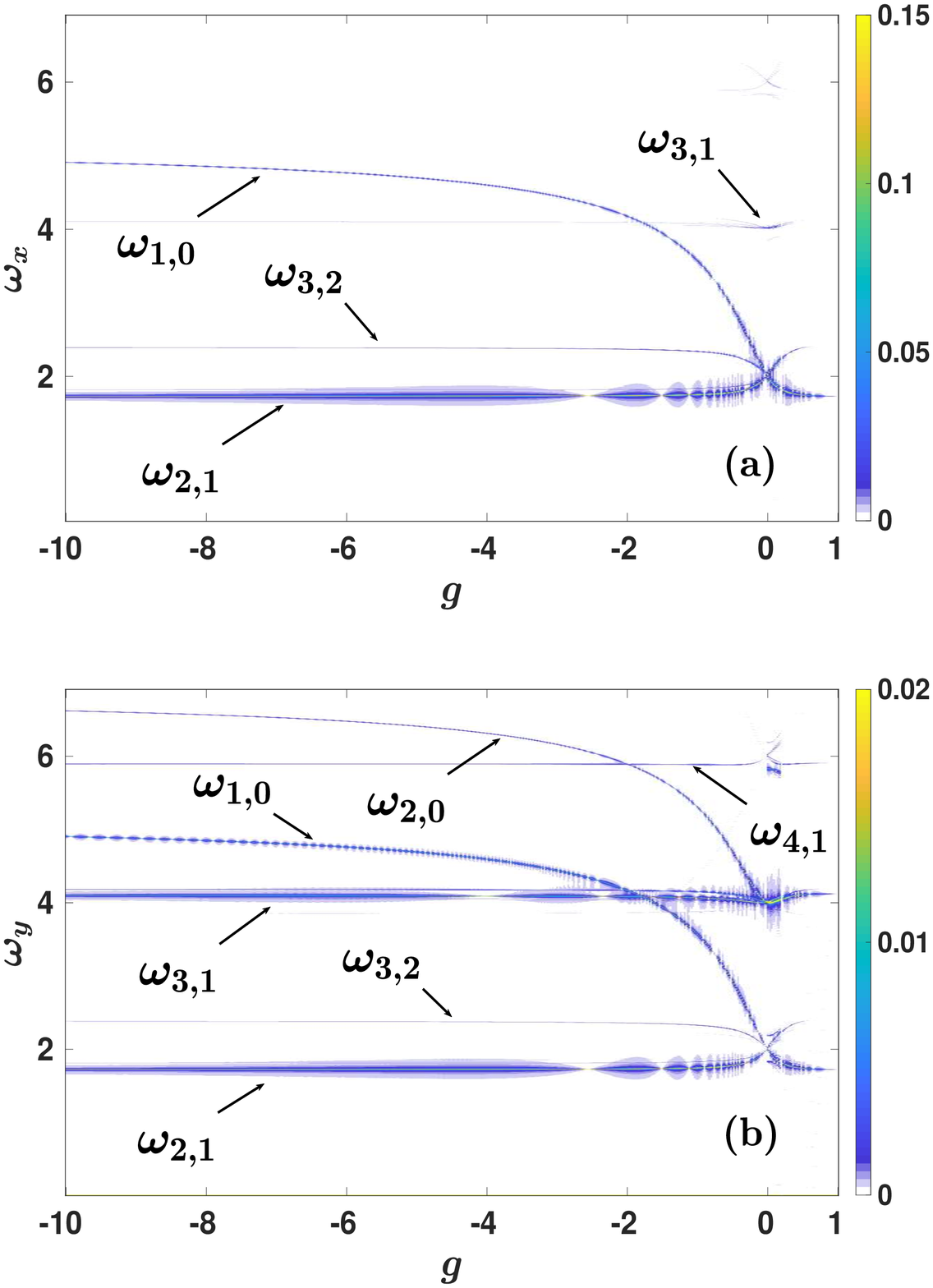}
\caption{Frequency spectrum (a) $F(\omega_x)$ of $\braket{x^2(t)}$ and (b) $F(\omega_y)$ of $\braket{y^2(t)}$. 
The anisotropy of the system is $\alpha=2$ and the interaction quench is performed from the ground state at $g^{\textrm{in}}=1$ 
to various attractive final interactions. 
The identified energy differences, $\omega_{ij}$, corresponding to the observed frequency branches are also shown. In all cases the quantities displayed are in dimensionless units.}
\label{Fig:Freqs_a_2}
\end{figure}

\begin{figure}[t!]
\centering
\includegraphics[width= 0.47 \textwidth]{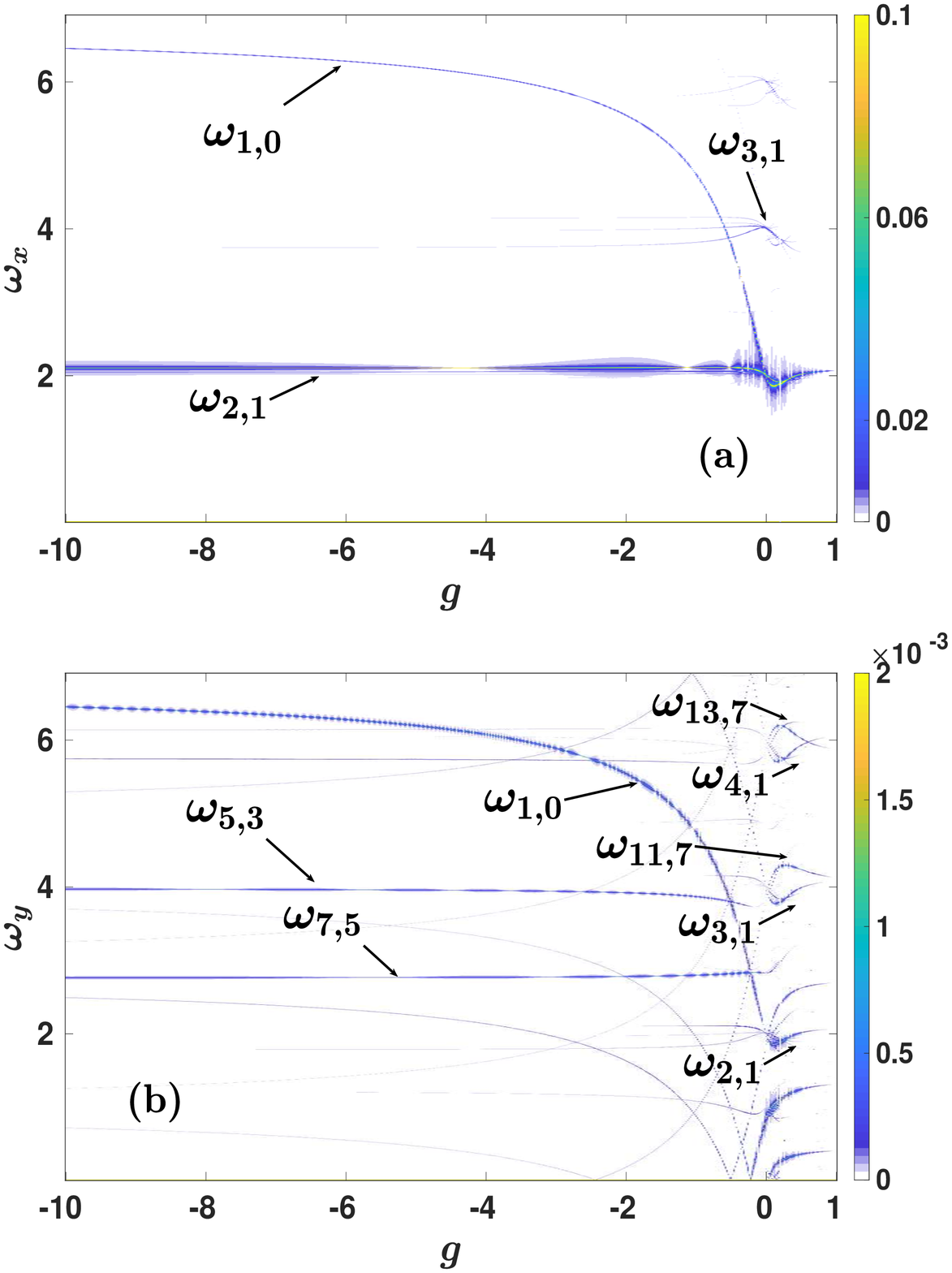}
\caption{Frequency spectrum (a) $F(\omega_x)$ of $\braket{x^2(t)}$ and (b) $F(\omega_y)$ of $\braket{y^2(t)}$. 
The anisotropy of the system is $\alpha=6.5$ following an interaction quench from the ground state at $g^{\textrm{in}}=1$ to different attractive 
final interactions. 
Specific energy differences, $\omega_{ij}$, referring to the observed frequency branches are also depicted. All quantities shown are in dimensionless units.}
\label{Fig:Freqs_a_6p5}
\end{figure}

Case examples of the above-mentioned frequency spectra are provided in Fig. \ref{Fig:Freqs_a_2} for $\alpha=2$ and in Fig. \ref{Fig:Freqs_a_6p5} for $\alpha=6.5$, upon applying an interaction quench 
from the ground state at $g^{\textrm{in}}=1$ towards the attractive interaction regime. 
Note that the emergent frequencies stem from the energy difference between specific eigenstates of the postquench Hamiltonian and will be denoted in the following as 
$\omega_{i,j}=\mathcal{E}_i-\mathcal{E}_j$ \cite{Laura,Bougas,Simos1}. 
Moreover the amplitude of these frequencies suggests their degree of participation in the time-evolution, which can be explicitly measured via the respective overlap coefficients [Eq. \ref{overlap_coef}]. 
The latter essentially means that a relatively large [small] amplitude of $\omega_{i,j}$ indicates a dominant [suppressed] contribution of the involved eigenstates. 
Regarding the motion of the bosons along the $x$ direction we calculate the frequency spectrum $F(\omega_x)$, see Fig. \ref{Fig:Freqs_a_2} (a). 
In the attractive interaction regime, there is a dominant frequency marked as $\omega_{2,1}$ which corresponds to the energy difference between the ground and the first excited state. 
Indeed by calculating the corresponding overlap coefficients [Eq. (\ref{overlap_coef})] for attractive postquench interactions, it turns out that the final ground state ($E_1$) possesses the largest population, while the next-to-leading-order occupied one is the first excited state ($E_2$).
Additionally, there are two other frequencies denoted by $\omega_{1,0}$ and $\omega_{3,2}$ possessing a relatively much smaller amplitude than $\omega_{2,1}$. 
These frequencies refer to the energy difference between the bound and the ground state and of the second with the first excited states respectively. 
Close to zero postquench interactions, all these frequencies approach $\omega_x\simeq 2$. 
The latter can be easily deduced by inspecting the corresponding energy spectrum at $\alpha=2$, see Fig. \ref{Fig:Spectra} (b), where the energy spacing is uniform at zero interactions 
in contrast to the non-uniform energy gaps appearing in both the repulsive and the attractive interaction regimes. 
Furthermore, in the vicinity of $g=0$ another frequency contributes to the spectrum of $\braket{x^2(t)}$, namely $\omega_{3,1}$ whose amplitude decreases substantially for attractive as well as repulsive interactions. 

Entering the repulsive interaction regime we observe that mainly two frequencies dominate, i.e. $\omega_{2,1}$ and $\omega_{3,2}$. 
Note that $\omega_{2,1}$ has a larger amplitude since it corresponds to the energy difference between the ground and the first excited state, which are the most significantly occupied states in this postquench interaction regime. 
Turning to the dynamical evolution in the $y$ direction the spectrum $F(\omega_y)$ is presented in Fig. \ref{Fig:Freqs_a_2} (b). 
Evidently, a larger number of frequencies are involved in the dynamics, but with an amplitude being of an order of magnitude smaller than the corresponding ones in the $x$ direction. 
The latter is attributed to the fact that the variance in the $y$ direction, which is tightly confined by the harmonic trap, is smaller compared to the one in the elongated $x$ direction. 
To qualitatively explain the larger number of frequencies along the $y$ direction one  can resort to an analytic expression for $F(\omega_y)$ and $F(\omega_x)$, namely
\begin{eqnarray}
    F(\omega_x)&=&\frac{B^2\sqrt{2\alpha}}{4\pi}\sum_{i,j} \delta[\omega_x-\omega_{i,j}]\mathcal{A}_x(i,j), \\
    F(\omega_y)&=&\frac{B^2\sqrt{2}}{4\alpha^{5/2}\pi}\sum_{i,j} \delta[\omega_y-\omega_{i,j}]\mathcal{A}_y(i,j).
    \label{More_Freq}
\end{eqnarray}
For the detailed derivation of these spectra as well as the explicit expressions of the involved amplitudes $\mathcal{A}_x(i,j)$ and $\mathcal{A}_y(i,j)$, see Appendix \ref{Ap:Freq}. 
It is worth mentioning here that both $\mathcal{A}_x(i,j)$ and $\mathcal{A}_y(i,j)$ depend on $\omega_{i,j}$. 
Closely comparing $\mathcal{A}_x(i,j)$ and $\mathcal{A}_y(i,j)$, see also Appendix \ref{Ap:Freq}, we can deduce that for $(i,j)=(1,2)$ $\mathcal{A}_x(i,j)\gtrsim \mathcal{A}_y(i,j)$ is satisfied, while for all other pairs $i\neq j>2$ it holds that $\mathcal{A}_y(i,j)> \mathcal{A}_x(i,j)$.  
The latter means that a larger number of frequencies contributes to $\mathcal{A}_y(i,j)$ than $\mathcal{A}_x(i,j)$ and especially the higher-order ones possess a vanishing contribution to $\mathcal{A}_x(i,j)$. 
In particular, for attractive interactions there are predominantly four contributing frequencies, namely $\omega_{2,1}$ and $\omega_{3,1}$, which stem from the energy difference between the ground and the first and second excited states respectively. 
Also, the frequencies $\omega_{1,0}$ and $\omega_{2,0}$ are imprinted in the spectrum and refer to the energy difference among the bound state and the ground or first excited state respectively. 
Approaching the non-interacting regime, $g=0$, two more frequencies appear i.e. $\omega_{4,1}$ and $\omega_{3,2}$ [hardly visible in Fig. \ref{Fig:Freqs_a_2} (b)]. 
Note that at $g=0$ all three frequencies, $\omega_{2,1}$, $\omega_{3,2}$ and $\omega_{1,0}$ merge to $\omega_y \simeq 2$, see also the previous discussion. 
However, on the repulsive regime essentially two frequencies dominate, i.e. $\omega_{2,1}$ and $\omega_{3,1}$. 

The frequency spectra of $\braket{x^2(t)}$ and $\braket{y^2(t)}$ for a larger anisotropy $\alpha=6.5$ and for the same interaction quench scenario as before 
are illustrated in Fig. \ref{Fig:Freqs_a_6p5}. 
Along the $x$ direction [Fig. \ref{Fig:Freqs_a_6p5} (a)] and for interparticle attractions, the most prominent frequency corresponds to the energy difference between the ground 
and the first excited state i.e. $\omega_{2,1}$. 
In terms of the involved overlap coefficients these two states have the dominant contribution during the dynamics. 
There is also another frequency, stemming from the energy difference of the bound and the ground state, $\omega_{1,0}$, which becomes more prominent close to zero postquench interactions. 
This frequency possesses a larger value compared to the corresponding one for $\alpha=2$, see also Fig. \ref{Fig:Freqs_a_2} (a), since the energy difference between the two involved states grows 
with increasing anisotropy parameter, as shown explicitly in Fig. \ref{Fig:energy_diff} (b). 
For $g\approx0$, there is an additional frequency present, namely $\omega_{3,1}$, which disappears for attractive as well as repulsive interactions. 
The frequencies regarding the dynamics along the $y$ direction [Fig. \ref{Fig:Freqs_a_6p5} (b)] are fainter from the respective ones in the $x$ direction by almost two orders of magnitude. 

Moreover for attractive interactions, more frequencies are involved in the dynamics in the strongly confined direction, with the most prominent one stemming from the energy difference between the 
ground and the bound state, $\omega_{1,0}$. 
In the vicinity of zero interactions, there is a multitude of frequencies referring to the energy difference between the ground and higher excited states such as $\omega_{2,1}$ and $\omega_{4,1}$, 
as well as frequencies stemming from higher-lying energy eigenstates e.g. $\omega_{11,7}$ and $\omega_{13,7}$. 
The larger number of frequencies in the $y$ direction and their smaller amplitude compared to the ones appearing along the $x$ direction can be explained with the same reasoning applied to Fig. \ref{Fig:Freqs_a_2} (b), see in particular the discussion in the context of Eq. (\ref{More_Freq}). 
Note here that some of the frequencies depicted in Fig. \ref{Fig:Freqs_a_6p5} (b) have a very small amplitude and are not identified by specific energy differences between the eigenstates 
of the system. 
A further increase of the anisotropy parameter $\alpha$, essentially freezes out the motion along the $y$ direction and the frequencies involved in the dynamics become fainter (not shown for brevity). 
The most prominent frequency that remains is the energy difference between the bound and the ground state in the attractive regime.

\subsection{One-body density evolution}

To unveil the dynamical spatial redistribution of the two bosons, subjected to an interaction quench, from a single-particle perspective we inspect their reduced one-body density which 
can be experimentally probed \cite{Sakmann,Jochim}. 
In particular, the time-evolution of the one-body reduced density starting from a state characterized by energy $\mathcal{E}_i^{\textrm{in}}$ at $g^{\textrm{in}}$ towards $g$ reads
\begin{widetext}
\begin{gather}
\rho^{(1)}(x_1,y_1;t)=\left( \frac{\sqrt{\alpha}}{\pi}\right)^3 e^{-(x_1^2+\alpha y_1^2)} \sum_{j,j'} e^{\iu (\mathcal{E}_j-\mathcal{E}_{j'})t} B_j B_{j'} d_{i,j}d^*_{i,j'} \sum_{n,m} \frac{H_n(0)H_m(0)}{2^{n+m+2}n!m!} \nonumber \\
\Gamma\left(\frac{\alpha n-\mathcal{E}_j}{2}\right) \Gamma\left(\frac{\alpha m-\mathcal{E}_{j'}}{2}\right) \int_{-\infty}^{\infty} dy_2 \, e^{-\alpha y_2^2} H_n\left(\sqrt{\alpha} \frac{y_1-y_2}{\sqrt{2}}\right) H_m\left(\sqrt{\alpha} \frac{y_1-y_2}{\sqrt{2}}\right) \nonumber \\
\int_{-\infty}^{\infty} dx_2 \, e^{-x_2^2} U\left(\frac{\alpha n-\mathcal{E}_j}{2},\frac{1}{2},\frac{(x_1-x_2)^2}{2} \right)U\left(\frac{\alpha m -\mathcal{E}_{j'}}{2},\frac{1}{2},\frac{(x_1-x_2)^2}{2}\right).
\end{gather}
\end{widetext}

\begin{figure}[t!]
	\centering
	\includegraphics[width=0.5 \textwidth]{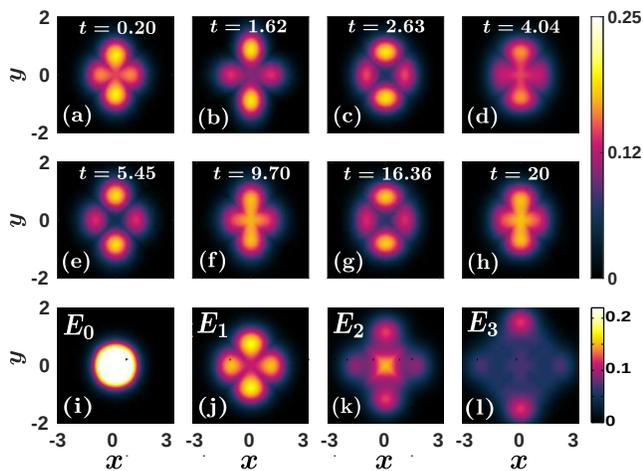}
	\caption{(a)-(h) Instantaneous one-body density following an interaction quench from the ground state at $g^{\textrm{in}}=1$ to $g=-0.2$. (i)-(l) One-body density of the dominantly populated postquench eigenstates in the time-evolution. 
		The system consists of two bosons and the anisotropy of the 2D harmonic trap is $\alpha=2$. For all observables dimensionless units are adopted.}
	\label{Fig:Density_a_2_evol}
\end{figure}

\begin{figure}[t!]
	\centering
	\includegraphics[width=0.5 \textwidth]{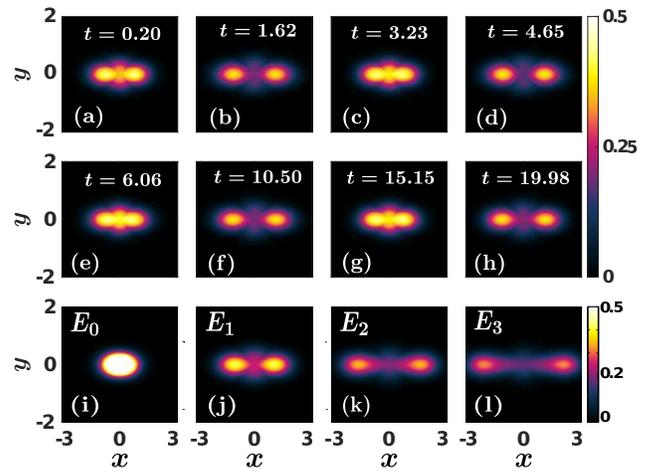}
	\caption{(a)-(h) Snapshots of the one-body density after an interaction quench from the ground state at $g^{\textrm{in}}=1$ to $g=-0.2$. (i)-(l) One-body density of the dominantly contributing postquench eigenstates during the dynamics. 
		The anisotropy of the 2D harmonic trap is $\alpha=6.5$. All quantities shown are in dimensionless units.}
	\label{Fig:Density_a_6p5_evol}
\end{figure}

Figures \ref{Fig:Density_a_2_evol} and \ref{Fig:Density_a_6p5_evol} display snapshots of the reduced one-body density for a quench from the ground state 
at $g^{\textrm{in}}=1$ to $g=-0.2$ for $\alpha=2$ and 6.5 respectively. 
We remark that the postquench interaction is close to the non-interacting regime where the time evolved state deviates significantly from the initial one, see also Fig. \ref{Fig:Mean_fid} (b). Also, the depicted time-instants correspond to the timescales set by the prevalent frequencies in the dynamics of the $x$ and $y$ direction variances identified in Figs. \ref{Fig:Freqs_a_2}, \ref{Fig:Freqs_a_6p5}. 
These frequencies are the energy differences between the predominantly contributing postquench eigenstates in the dynamics of the system as it can also be verified by calculating the respective overlap coefficients [Eq. (\ref{overlap_coef})]. 

Referring to the case of $\alpha=2$ [Fig. \ref{Fig:Density_a_2_evol}] we observe the appearance of two-humped structures in both the $x$ and $y$ directions, see for 
instance Figs. \ref{Fig:Density_a_2_evol} (a), (b), (c) and (e). 
The appearance of these hump patterns is predominantly attributed to the participation of the postquench eigenstates, $E_1$ [Fig. \ref{Fig:Freqs_a_2} (j)] and $E_2$ [Fig. \ref{Fig:Freqs_a_2} (k)] during the dynamics. 
Notably the eigenstate with energy $E_2$ has a relatively much smaller impact on the shape of $\rho^{(1)}(x_1,y_1;t)$ compared to one with energy $E_1$, a result that is also confirmed by inspecting the corresponding overlap coefficients since $d_{1,1}\gg d_{1,2}$. 
However, during the contraction of the bosonic cloud, the two-hump structure is destroyed by means of a smoothening of the density profile and the development of a cross-like 
pattern [Figs. \ref{Fig:Density_a_2_evol} (f), (h)]. 
This structural change of $\rho^{(1)}(x_1,y_1;t)$ is caused by the predominant contribution of the  postquench bound state with energy $E_0$ [Fig. \ref{Fig:Freqs_a_2} (i)], whose presence is manifested in the contraction of the cloud.
Note that the contraction of the bosons is identified by inspecting the time-evolution of $\braket{x^2(t)}$ and $\braket{y^2(t)}$ (not shown for brevity). 
In particular, when $\braket{x^2(t)}$ and $\braket{y^2(t)}$ experience minima [maxima] the bosons feature a contraction [expansion]. 
Moreover, the two-hump structure shown in the one-body density [Figs. \ref{Fig:Density_a_2_evol} (b), (c) and (e)] is associated with the expansion of the cloud, a result that can 
again been confirmed from the dynamics of $\braket{x^2(t)}$ and $\braket{y^2(t)}$. 

For a larger anisotropy, e.g. $\alpha=6.5$ shown in Fig. \ref{Fig:Density_a_6p5_evol}, the motion along the $y$ direction is frozen out, as anticipated by the frequency spectra presented 
in Fig. \ref{Fig:Freqs_a_6p5} (b). 
Thus, the single-particle density evolution takes place predominantly along the $x$ direction, and corresponds to a breathing dynamics. 
Indeed, when the density expands there is a two-hump structure, see Figs. \ref{Fig:Density_a_6p5_evol} (b), (d) and (f), whilst for a contraction in the $x$ direction 
[see Figs. \ref{Fig:Density_a_6p5_evol} (c), (e) and (g)] the two-hump pattern disappears and the density dip around the trap center is filled. 
Again, the contraction and expansion of the two bosons is identified by inspecting the minima and maxima of $\braket{x^2(t)}$ and $\braket{y^2(t)}$ after the quench. 
We finally remark that the time-evolved state resides mainly in a superposition of the bound, $E_0$ [Fig. \ref{Fig:Density_a_6p5_evol} (i)], and the ground state, $E_1$ [Fig. \ref{Fig:Density_a_6p5_evol} (j)]. 
This fact is verified by calculating the corresponding overlap coeffcients [Eq. (\ref{overlap_coef})] and it is also readily supported by comparing the instantaneous $\rho^{(1)}(x_1,y_1;t)$ with the $\rho^{(1)}(x_1,y_1;0)$ of the corresponding postquench eigenstates. 
Other energetically higher-lying excited states have a much smaller contribution in the time-evolved two-body state and thus their impact is less obvious in $\rho^{(1)}(x_1,y_1;t)$, see e.g. Figs. \ref{Fig:Density_a_6p5_evol} (k), (l) for $E_2$ and $E_3$ respectively. 

\section{Summary and outlook} \label{conclusion}

We have investigated the stationary properties and the interaction quench dynamics of two bosons confined in an anisotropic 2D harmonic trap, 
and interacting through an $s$-wave pseudo-potential. 
A transcendental equation with respect to the anisotropy parameter is derived giving access to the energy spectrum of the system. 
The spectrum is in turn explored for a wide range of attractive and repulsive 2D coupling strengths and arbitrary values of the anisotropy.  

It is found that the energy spacing between the involved energy eigenstates for a fixed interaction strength strongly depends on the anisotropy. 
Deep into the quasi-1D regime, where the anisotropy is very large, the energy spectrum of the purely 1D setup is retrieved. 
Importantly, a relation is established between the two- and the 1D scattering lengths. 
Moreover, we have derived an analytical expression for the two-boson wavefunction both in real and momentum space. 
It is shown that for interparticle distances much smaller than the harmonic oscillator length in the less tightly confined direction the wavefunction exhibits a logarithmic 
singularity, a feature which is inherently related to two spatial dimensions. 
In momentum space, the wavefunction exhibits a multihump structure along the weaker confined direction with the humps being elongated along the other direction. 
This latter behavior becomes more prominent as the anisotropy increases. 
The corresponding one-body densities feature a two-hump structure along the spatial direction where the confinement is less tight, a behavior that is more pronounced 
for a larger anisotropy. 
For higher-lying excited states the inter-hump separation is enhanced. 

Subsequently we have investigated the Tan contact, which captures short-range two-body correlations, for different anisotropies in both the repulsive and the 
attractive interaction regimes. 
Inspecting the contact of the bound state reveals an increasing tendency for larger anisotropies independently of the sign of the interaction and does not saturate as 
the quasi-1D region is approached. 
Furthermore, the short-range two-body correlations of the ground state increases for small anisotropies and subsequently saturates for larger ones. 
Within the quasi-1D regime, a relation is established among the two- and the 1D contacts unveiling that they are proportional by a geometric factor and the 
harmonic oscillator length along the strongly confined direction. 

Apart from the stationary properties, we have also examined the dynamical evolution of the system by applying an interaction quench for different anisotropies. 
Employing the time-averaged fidelity of the system we have showcased that the time-evolved state deviates significantly from the initial one in the vicinity of zero postquench interactions whilst it 
is less perturbed for stronger postquench interactions. 
Moreover, for increasing anisotropy the system becomes less perturbed after an interaction quench of fixed amplitude in both the attractive and the repulsive coupling regimes. 
The quench excites a breathing motion in both the $x$ and the $y$ directions, with a distinct number of participating frequencies in each spatial direction. 
At large anisotropies the motion along the $y$ direction freezes out, and there are many eigenstates contributing in the dynamics with the most prominent one being 
the bound state. 
The dynamical response is also visualized on the one-body level, by monitoring the evolution of the reduced one-body density after an interaction quench in the vicinity 
of zero interactions, where the time-evolved state deviates substantially from the initial one. 
For small anisotropies the bosonic cloud undergoes a periodic expansion and contraction dynamics in both spatial directions, with the appearance of a two hump structure building upon the 
one-body density in both the $x$ and $y$ spatial directions. 
An increasing anisotropy, causes density oscillations and the development of two humps along the less tight direction, while the motion in the tightly confined direction is frozen out. 

There are several research directions that one can pursue in future works. 
A straightforward extension is to perform a quench of the anisotropy parameter, and investigate the resulting non-equilibrium dynamics of the two-bosons from the 2D plane to 
the quasi-1D regime and vice versa. 
Here, it is interesting to inspect how efficiently one can populate specific eigenstates since this quench changes the energy gaps between the various states. 
Another prospect is to consider a long-range interaction between the atoms, such as a dipolar coupling, in order to study how the long-range character affects the energy spectra and 
also the non-equilibrium dynamics. 
Finally, the extension to three interacting bosons in an anisotropic 2D trap and exploring their stationary and dynamical properties is certainly of interest. 
The latter endeavor can shed light e.g. into the dynamical formation of trimer bound states. 

\begin{acknowledgments}
G. B. and S. I. M. would like to thank A. I. Karanikas for fruitful discussions regarding the integral formulas. 
G. B. kindly acknowledges financial support by the State Graduate Funding Program Scholarships (HmbNFG). 
S. I. M.  gratefully  acknowledges financial support in the framework of the Lenz-Ising Award of the Department 
of Physics of the University of Hamburg. This work was supported within the framework of the PIER Hamburg-MIT
Germany program funded  by the Ministry of Science, Research and
Equalities of the Free and Hanseatic City of Hamburg. 
P. S. gratefully acknowledges financial support by the Deutsche Forschungsgemeinschaft (DFG) in the framework of the 
SFB 925 ``Light induced dynamics and control of correlated quantum systems" (DFG 170620586). 
\end{acknowledgments} 

\appendix 

\section{Transcendental equation for the relative energies} \label{Trans} 

In this appendix the transcendental equation for determining the energy of two bosons confined in a 2D harmonic trap with anisotropy parameter $\alpha$ is derived. 
Plugging Eq. \eqref{integral_split} into Eq. \eqref{energy} and performing the change of variables $z=e^{-t}$ in $I(f(E)/2)$, the equation that determines the energy of the 
system reads 
\begin{equation}
-\gamma +\ln L+\ln 2 +\sqrt{\alpha}\int_{0}^{e^{-L}} dz\, \frac{z^{f(E)/2-1}}{\sqrt{1-z}\sqrt{1-z^\alpha}}=\ln (a_{\textrm{2D}}^2).
\label{energy_spectrum}
\end{equation} 
As it has already been remarked in Sec. \ref{Setup}, the integral appearing in the general form of the wavefunction [Eq. (\ref{wavefunction_integral})] 
converges for $f(E)>0$, which corresponds to eigenstates with energy lower than $\frac{\alpha+1}{2}$. 
To extend Eq. (\ref{energy_spectrum}) to energies larger than the zero point oscillation energy, we shall use the following relation that the 
integral $I(f(E)/2)$ satisfies:
\begin{equation}
I(f(E)/2)=I(\alpha+f(E)/2)+\int_{0}^{e^{-L}}dz \, \frac{z^{f(E)/2-1}\sqrt{1-z^{\alpha}}}{\sqrt{1-z}}.
\end{equation}
The latter integral can be performed analytically, if the term $\sqrt{1-z^{\alpha}}$ is expanded as a Taylor series yielding
\begin{eqnarray}
I\left(\frac{f(E)}{2}\right)&=&I\left(\alpha+\frac{f(E)}{2}\right)+ \nonumber \\& & \sum_{n=0}^{\infty} \binom{1/2}{n}\frac{\sqrt{\pi}(-1)^n\Gamma\left(
\frac{f(E)}{2}+\alpha n\right)}{\Gamma\left(\frac{1}{2}+\frac{f(E)}{2}+\alpha n\right)}. \nonumber \\
\label{extension}
\end{eqnarray}
The last point that one needs to take care of is the divergence of the integral $I(f(E)/2)$ as $L\rightarrow 0$. 
This divergence turns out to be logarithmic and it can be extracted from the following integral
\begin{equation}
I\left(\frac{f(E)}{2}\right)=-\frac{\ln L}{\sqrt{\alpha}}+\int_{0}^{1} dz \, \ln(1-z)\varphi'\left(z,\frac{f(E)}{2}\right),
\end{equation}
where $\varphi\left(z,\frac{f(E)}{2}\right)=z^{f(E)/2-1}\frac{\sqrt{1-z}}{\sqrt{1-z^{\alpha}}}$ and the differentiation is with respect to the variable $z$. 
Moreover, the first term cancels exactly the term $\ln L$ present in the transcendental Eq. \eqref{energy_spectrum}. 
We can further express Eq. \eqref{energy_spectrum} in the form
\begin{equation}
-\gamma +2\ln 2+\sqrt{\alpha}\int_{0}^{1} dz \, \ln(1-z)\varphi'\left(z,\frac{f(E)}{2}\right)=-\frac{1}{g}.
\label{energy_spectrum_1}
\end{equation}
The latter is exactly the transcendental equation that we are seeking. 
We remark that Eq. \eqref{extension} extends the validity of Eq. \eqref{energy_spectrum_1} to $f(E)<0$, determining thus completely 
the relative energy of the two bosons. 

\section{Retrieving the 1D spectrum} \label{Recovery} 

To recover the well-known 1D energy spectrum from the transcendental Eq. (\ref{energy_spectrum_1}) we assume that $\alpha \gg 1$. 
In this case one can separate the integral $I\left(\frac{f(E)}{2}\right)$ into two parts, namely 
\begin{eqnarray}
I\left(\frac{f(E)}{2}\right)&=& \overbrace{\int_{0}^{\theta} dx\, \frac{x^{f(E)/2-1}}{\sqrt{1-x}}}^{I_1} \nonumber \\ & &+\underbrace{\int_{\theta}^{e^{-L}} dx\, \frac{1}{\sqrt{1-x}\sqrt{1-x^{\alpha}}}}_{I_2},
\end{eqnarray}
where $\theta$ is a parameter very close to unity, such that $\frac{1}{\sqrt{1-x^{\alpha}}}\leq 1+\epsilon$ on the interval $[0,\theta]$, with $\epsilon \ll 1$. 
In this case, $\theta=1-\frac{k}{\alpha}$, where $k\approx 6$ for achieving an accuracy of $\epsilon \approx 0.001$. 
Therefore, $I_1$ reads  
\begin{equation}
  I_1=\sqrt{\pi} \frac{\Gamma\left( \frac{f(E)}{2}  \right)}{\Gamma\left( \frac{1}{2}+\frac{f(E)}{2} \right)}-2\sqrt{\frac{k}{\alpha}} +\mathcal{O}(\alpha^{-3/2})
  \label{I_1}
\end{equation}
assuming that $\theta$ is very close to 1. 
In the second part, $I_2$, the dependence on the energy is dropped, since in this interval $x$ is very close to unity. 
Furthermore, the term $1/\sqrt{1-x^{\alpha}}$ can be expanded for $x$ close to unity as follows 
\begin{gather}
     \frac{1}{\sqrt{1-x^{\alpha}}}=\frac{1}{\sqrt{{\alpha}}\sqrt{1-x}}+\frac{(\alpha-1)\sqrt{1-x}}{4\sqrt{\alpha}} \nonumber \\
    +\frac{(\alpha^2+6\alpha-7)(1-x)^{3/2}}{96\sqrt{\alpha}} \nonumber \\  +\frac{(\alpha^3-3\alpha^2-13\alpha+15)(1-x)^{5/2}}{384\sqrt{\alpha}} \nonumber \\ 
    +\mathcal{O}\left(\frac{(1-x)^{7/2}}{10240} \right).
\end{gather}
Keeping the first four terms, the integral $I_2$ becomes
\begin{gather}
    I_2=-\frac{\ln L}{\sqrt{\alpha}}+\frac{\ln(k/\alpha)}{\sqrt{a}}+\frac{k}{4\sqrt{\alpha}}-\frac{k^2}{192\sqrt{\alpha}} \nonumber \\ -\frac{k^3}{1152\sqrt{\alpha}} 
    +\mathcal{O}\left( \alpha^{-1} \right)
    \label{I_2}
\end{gather}
The other terms are of the order of $\mathcal{O}\left(\frac{1}{\alpha}\right)$ and for sufficiently large $\alpha$ become negligible. 
Gathering the two integrals $I_1$ and $I_2$ [Eqs. \eqref{I_1}, \eqref{I_2}] together, the transcendental Eq. \eqref{energy_spectrum} becomes
\begin{gather}
  -\gamma+\sqrt{\pi \alpha} \frac{\Gamma\left(\frac{f(E)}{2}\right)}{\Gamma\left(\frac{1}{2}+\frac{f(E)}{2}\right)}-2\sqrt{k}+\ln (2k) -\ln (\alpha)+\frac{k}{4}\nonumber \\
  -\frac{k^2}{192}- 
  -\frac{k^3}{1152}=\ln(a_{\textrm{2D}}^2)
\end{gather}
This expression is the transcendental equation of two bosons deep into the quasi-1D regime.

\section{The Tan contact and its quasi-1D limit} \label{Ap:Tan}

To find the Tan contact, we start from the 2D Fourier transform of a radially symmetric wavefunction $\Psi(\rho)$ \cite{Bougas} namely  
\begin{equation}
\tilde{\Psi}(k,t)= 2\pi \int_{0}^{\infty} d\rho\, \rho \Psi(\rho,t) J_0(2\pi\rho k),
\label{2D-Fourier}
\end{equation}
where $J_0(x)$ denotes the zeroth order Bessel function. 
In our setup, the wavefunction $\Psi(x,y)$ is radially symmetric only for small $x,y$. 
Thus, if we restrict the integration at very small values of $\rho$, i.e. very large momenta, the contact is obtained from the leading order term ($\sim 1/k^2$) 
in the resulting expression \cite{Bougas}, and reads 
\begin{equation}
\mathcal{D}(\alpha,\mathcal{E})=\frac{B^2(\alpha,\mathcal{E})}{4\pi^4}.
\label{stat_contact}
\end{equation}
Moreover, if $\alpha=1$, Eq. \eqref{stat_contact} reduces to $\mathcal{D}(1,\mathcal{E})=\frac{1}{\pi^3\psi^{(1)}(-\mathcal{E}/2)}$, which is the contact of a stationary eigenstate in an 
isotropic 2D trap \cite{Bougas}, and $\psi^{(1)}(z)$ is the trigamma function \cite{Stegun}.

For large $\alpha$, i.e. in the quasi-1D regime, only the term $m=0$ dominates in the summation of Eq. \eqref{norma} for the normalization constant $B$. 
Hence, in this case the contact can be written as follows 
\begin{equation}
\frac{B^2(\alpha \gg 1,\mathcal{E})}{4\pi^4}=\frac{1}{\pi^{7/2}}\frac{\Gamma\left(-\frac{\mathcal{E}}{2}+\frac{1}{2} \right)}{\Gamma\left(-\frac{\mathcal{E}}{2} \right) \left[
\psi\left(\frac{1-\mathcal{E}}{2} \right)-\psi\left(-\frac{\mathcal{E}}{2}\right) \right]\sqrt{\alpha}}.
\label{Contact_limit}
\end{equation}
This form is analogous to the Tan contact for two interacting bosons confined in a 1D harmonic trap \cite{Vignolo,Corson}, rescaled by the anisotropy parameter $\alpha$. 
To be more precise, the 1D Tan contact, when adopting the same convention for the Fourier transform as in Eq. \eqref{2D-Fourier}, 
namely $\tilde{\Psi}(k)=\int_{-\infty}^{\infty} dx\, e^{-2\pi \iu k x} \Psi(x)$, reads \cite{Vignolo}
\begin{equation}
\mathcal{D}_{\textrm{1D}}=\frac{\Gamma\left(\frac{1}{2}-\epsilon\right)}{\pi^4\Gamma(-\epsilon)\left[ \psi\left(\frac{1}{2}-\epsilon\right)-\psi\left(-\epsilon \right) \right]},
\end{equation}
where $\epsilon=\frac{E}{2}-\frac{1}{4}$, and the energy $E$ is determined by the transcendental Eq. \eqref{spectrum_1D}. When restoring the units of the system, a relation is established among the 1D and the 2D contacts namely
\begin{equation}
\mathcal{D_{\textrm{2D}}}=l_y \sqrt{\pi}\mathcal{D}_{\textrm{1D}},
\end{equation} 
which holds in the quasi-1D regime.

\section{Analytical expression for the frequency amplitudes of the two atom variance} \label{Ap:Freq}

The frequency amplitudes of the spatial extent of the two atoms during the dynamics can be analytically determined, by employing the following expansion of the time-evolved relative wavefunction in terms of the postquench eigenstates
\begin{equation}
    \Psi_{\textrm{rel},i}^{\textrm{in}}(x,y;t)=\sum_j e^{-\iu \mathcal{E}_j t} \Psi_{\textrm{rel},j}^f(x,y)d_{i,j}. 
\end{equation}
Here $\Psi_{\textrm{rel},j}^f(x,y)$ are the postquench eigenstates (see also Eq. \eqref{Waves_simpl}) with energy $\mathcal{E}_j=E_j-(\alpha+1)/2$. 
Also, $d_{i,j}$ denote the overlap coefficients between the postquench and initial eigenstates [Eq. (\ref{overlap_coef})].
Substituting the above relation into $\braket{x^2(t)}=\int_{-\infty}^{\infty} dxdy \, x^2 \abs{\Psi^{\textrm{in}}_{\textrm{rel},i}(x,y;t)}^2$ and performing the integration over the $y$ direction, we obtain
\begin{eqnarray}
\braket{x^2(t)}&=&\frac{B^2\sqrt{\alpha}}{4\pi^{3/2}}\sum_{j,j'}d_{i,j}d_{i,j'}e^{-i(\mathcal{E}_j-\mathcal{E}_{j'})t}\sum_m\frac{H_m^2(0)}{2^m m!} \nonumber \\
& &\Gamma\left( \frac{\alpha m-\mathcal{E}_j}{2} \right)\Gamma \left( \frac{\alpha m-\mathcal{E}_{j'}}{2} \right) I^m_{j,j'}, 
\end{eqnarray} 
where the last integral reads 
\begin{eqnarray}
I^m_{j,j'}&=&\int_{-\infty}^{\infty} dx \, x^2 e^{-x^2}U\left(\frac{\alpha m-\mathcal{E}_j}{2},\frac{1}{2},x^2 \right) \nonumber \\
 & &\times U\left(\frac{\alpha m-\mathcal{E}_{j'}}{2},\frac{1}{2},x^2 \right).
\label{variancex}
\end{eqnarray}

Along the same lines we can calculate the explicit expression for $\braket{y^2(t)}$, namely 
\begin{eqnarray}
    \braket{y^2(t)}&=&\frac{B^2}{4\alpha^{5/2}\pi^{3/2}}\sum_{j,j'}d_{i,j}d_{i,j'}e^{-i(\mathcal{E}_j-\mathcal{E}_{j'})t}\sum_n\frac{H_n^2(0)}{2^n n!} \nonumber \\
& &\Gamma\left( \frac{ n-\mathcal{E}_j}{2\alpha} \right)\Gamma \left( \frac{ n-\mathcal{E}_{j'}}{2\alpha} \right) I^n_{j,j'},
\end{eqnarray}
with the latter integral having the form
\begin{eqnarray}
I^n_{j,j'}&=&\int_{-\infty}^{\infty} dy \, y^2 e^{-y^2}U\left(\frac{ n-\mathcal{E}_j}{2\alpha},\frac{1}{2},y^2 \right) \nonumber \\
 & &\times U\left(\frac{ n-\mathcal{E}_{j'}}{2\alpha},\frac{1}{2},y^2 \right).
\label{variancey}
\end{eqnarray}

Taking the Fourier transform of both $\braket{x^2(t)}$ and $\braket{y^2(t)}$, we find
\begin{eqnarray}
F(\omega_x)&=&\frac{B^2\sqrt{2\alpha}}{4\pi}\sum_{j,j'}\delta[\omega_x-\omega_{j,j'}]\mathcal{A}_x(j,j'), \\
F(\omega_y)&=&\frac{B^2\sqrt{2}}{4\alpha^{5/2}\pi}\sum_{j,j'}\delta[\omega_y-\omega_{j,j'}]\mathcal{A}_y(j,j'),
\end{eqnarray}
where the energy differences between the initial and the postquench eigenstates are $\omega_{j,j'}=\mathcal{E}_j-\mathcal{E}_{j'}$. 
Importantly, the corresponding amplitudes in the $x$ and $y$ spatial directions read 
\begin{eqnarray}
\mathcal{A}_x(j,j')&=&d_{i,j}d_{i,j'}\sum_m \frac{H_m^2(0)}{2^m m!}\Gamma\left( \frac{\alpha m-\mathcal{E}_j}{2} \right)\nonumber \\
& &\times \Gamma \left( \frac{\alpha m-\mathcal{E}_{j'}}{2} \right)I^m_{j,j'} \\
\mathcal{A}_y(j,j')&=&d_{i,j}d_{i,j'}\sum_n \frac{H_n^2(0)}{2^n n!}\Gamma\left( \frac{ n-\mathcal{E}_j}{2\alpha} \right)\nonumber \\
& &\times \Gamma \left( \frac{ n-\mathcal{E}_{j'}}{2\alpha} \right)I^n_{j,j'}.
\end{eqnarray} 
Inspecting these amplitudes for fixed $j$, $j'$ we can conclude by a direct numerical evaluation that for $j'=j+1$ and $j=1,2$, i.e. the ground and the first excited states, it holds that $\mathcal{A}_x(j,j')\gtrsim \mathcal{A}_y(j,j')$. 
Otherwise, it is found that $\mathcal{A}_y(j,j')> \mathcal{A}_x(j,j')$. 
As a consequence, in this latter case, there is a larger number of participating frequencies in $F(\omega_y)$ than $F(\omega_x)$ and therefore in the dynamics of the $y$ spatial direction. 
Indeed, by calculating numerically $\mathcal{A}_y(j,j')$ [$\mathcal{A}_x(j,j')$] it can be shown that higher-lying energy states possess a non-negligible [suppressed] contribution.

\end{document}